\documentstyle[12pt,epsfig]{article}
\newcommand{\gsim}{\lower.7ex\hbox{$\;\stackrel{\textstyle>}{\sim}\;$}}
\newcommand{\lsim}{\lower.7ex\hbox{$\;\stackrel{\textstyle<}{\sim}\;$}}
\newcommand{\scr}{\scriptscriptstyle}
\def\lam{\lambda}
\def\eps{\epsilon}
\def\bm#1{{\mbox{\boldmath $#1$}}}
\def\tc{t_{\rm c}}
\def\rc{R_{\rm c}}
\def\ri{R_{\rm i}}
\def\muc{\mu_{\rm c}}
\def\se{S_{\rm E}}
\def\te{t_{\rm E}}
\def\arccoth{{\rm arccoth}}
\def\sgn{{\rm sgn}}
\newcommand{\li}[1]{\,{\rm Li}_{#1}}
\def\umi{u_{-}}
\def\upl{u_{+}}
\def\vmi{{v_{-}}}
\def\vpl{{v_{+}}}
\def\tht{\theta_{\rm t}}
\def\tha{\theta_{\rm a}}
\def\sech{{\rm sech}}
\begin{document}
\footnotesep=14pt
\begin{flushright}
\baselineskip=14pt
{\normalsize DAMTP-1999-62}\\
{\normalsize {hep-ph/9907437}}
\end{flushright}
\renewcommand{\thefootnote}{\alph{footnote}}
\baselineskip=24pt

\vspace*{.5cm}
\begin{center}
{\Large\bf
Phase Equilibration and Magnetic Field}\\
{\Large\bf
Generation in U(1) Bubble Collisions}
\baselineskip=16pt
\vspace{0.75cm}

{\bf E.\,J. Copeland\footnote{Electronic address:
{\tt e.j.copeland@sussex.ac.uk}},
P.\,M. Saffin\footnote{ Electronic
address: {\tt p.m.saffin@damtp.cam.ac.uk}} and
O. T\"{o}rnkvist\footnote{
Electronic address: {\tt o.tornkvist@damtp.cam.ac.uk}}}\\
\vspace{0.4cm}
\footnotemark[1]{\em Centre
for Theoretical Physics, University of Sussex},\\
{\em Brighton BN1 9QH, United Kingdom}\\
\vspace{0.3cm}
\footnotemark[2]$^{,}$\footnotemark[3]{\em Department of Applied
Mathematics and Theoretical Physics\/},\\
{\em University of
Cambridge\/},\\{\em Cambridge CB3~9EW, United Kingdom}\vspace*{0.75cm}\\

{July 17, 1999}\\
\end{center}
\baselineskip=20pt
\begin{quote}
\begin{center}
{\bf\large Abstract}
\end{center}
\vspace{0.2cm}
{\baselineskip=10pt
We present the results of lattice computations of collisions of
two expanding bubbles of true vacuum
in the Abelian Higgs model with a first-order
phase transition.
New time-dependent
analytical solutions for the Abelian field strength and
the phase of the complex field are derived from initial conditions
inferred from linear superposition and are shown to be in excellent
agreement with the numerical solutions especially for the case where
the initial phase difference
between the bubbles is small.
With a step-function approximation for the
initial phase of the complex field,
solutions for
the Abelian field strength and other gauge-invariant quantities
are obtained in closed
form.
Possible extensions of the solution to the case of the electroweak
phase transition and the generation of primordial magnetic fields
are briefly discussed. }
\vspace*{8pt}
\noindent

\end{quote}
\renewcommand{\thefootnote}{\arabic{footnote}}
\setcounter{footnote}{0}
\newpage
\baselineskip=16pt

\section{Introduction}

Phase transitions are a universal feature
in
field
theories.
For example, they may
explain how quark confinement
has
emerged and possibly
how the electroweak symmetry was broken.
It is also through such transitions that
topological defects
form in condensed-matter systems in the laboratory and,
we believe, could have formed
in the early universe
(see \cite{vilshel} and
\cite{vachvol} for reviews).
Phase
transitions
of the first order
play
a particularly significant role in particle
cosmology,
either as a
progenitor of inflation \cite{guth,steinhardt,lyth} or as a source of the
baryon asymmetry associated with the electroweak transition \cite{liu}.
Popular
models of electroweak physics
such as the Minimal Supersymmetric Standard
Model exhibit a first-order transition for a range of experimentally
allowed parameters \cite{LaineCline}.
First-order phase transitions are
driven by the nucleation
of bubbles of true vacuum within
a sea of false vacuum. The bubbles,
responding to the pressure caused
by the energy difference of the two vacua, expand and subsequently collide.
Over the years, a number of authors have turned their attention to
the dynamics of such bubble collisions in gauge theories
\cite{Hawking,rudaz,Davis,KibVil,CopSaf,SafCop}.
In spite of this effort,
it is fair to say that much remains
unclear
about the role the various fields play in the collision process. This
should not
be surprising, as we are here dealing with a non-linear system which
is difficult to treat analytically.

Much of the interest in bubble collisions derives from the
pioneering work
of Kibble \cite{kibble}, who demonstrated
the possible formation of topological defects in a cosmological
phase transition for cases
where the vacuum manifold of the Higgs potential
is non-trivial.
In particular,
defects such as strings (or monopoles)
may be produced in the collision of three (four) bubbles, provided that inside
each bubble
the scalar Higgs field takes a different random value on the vacuum manifold.
In Ref.~\cite{kibble} the mechanism for defect formation was formulated
in terms of
a geodesic rule,
according to which the
phase (or, more generally, the isospace orientation)
of the
scalar
field
between two colliding bubbles
takes values
along the shortest path
on the vacuum manifold that connects the values
in
each bubble.
The geodesic rule has since been verified for defect formation in various
models with broken global symmetry \cite{globexp,sriv}.
In gauge
theories,
however, Rudaz and
Srivastava \cite{rudaz}
have argued that the geodesic rule does not apply,
since the phase of the scalar field has no
gauge-invariant meaning.
Instead, they suggested that
the dynamics of the fields would
play an important role in the production of gauge defects
in a first-order phase transition
and could
possibly lead to
a suppression in the rate of defect formation.

This claim was subsequently investigated by Hindmarsh, Davis,
and Brandenberger \cite{Davis}. Assuming
the gauge fields to be negligible at the moment of
collision, they obtained solutions which appeared to adhere to the
prescription of the (naive) geodesic rule
and found no suppression in
the probability of defect formation.
Later, however,
Copeland and Saffin \cite{CopSaf} demonstrated unequivocally
that the geodesic
rule can be violated in a U(1) gauge theory
as a consequence of the detailed energetics in bubble
collisions.
The reason is that
violent fluctuations in the modulus of the scalar field
cause symmetries to be restored locally,
allowing the phase to
``slip'' by an integer multiple of $2\pi$ and thus change the number
of string defects.
In fact,
when
the initial phase difference of the bubbles
and their initial separation are sufficiently large,
string defects can arise when only two bubbles collide,
unlike the usual picture of string formation obeying the geodesic
rule, which would require at least three bubbles.
In this
article, however, we will be dealing with phase differences and
bubble separations which are too small to cause defects to form
through violation of the geodesic rule.

An appealing aspect of
first-order transitions
involving
gauge fields is that
the
bubble collisions give rise to magnetic fields.
These are produced by a gauge-invariant current which develops
across the surface of intersection of the two bubbles due to the gradient
of the phase of the scalar field \cite{KibVil}.
Because bubble collisions occur
out of thermal equilibrium and the bubble walls possess large
kinetic energy, the magnetic fields thus created can be considerably
larger than the thermal background of magnetic-field fluctuations. If the
energetics permits the geodesic rule to be violated,
these magnetic fields become associated with the formation of
local (gauged)
defects.

Magnetic fields could also be generated
in extensions of the SU(2)$\times$U(1)
Standard Model
with a first-order phase transition and have been studied by a number of
authors
\cite{enqvist,OTmag1,OTmag2}.
Such fields,
created in the early universe,
could evolve into seed fields for a galactic dynamo mechanism
that would in turn amplify them to the strength observed in galaxies today.
The common
view has been that it is generally
too difficult to produce large enough fields
at the electroweak phase transition on sufficiently large
length scales, but recently Davis, Lilley and T\"o{}rnkvist \cite{Davis1}
have argued that in the context of a universe with less matter than
the critical
density (and particularly with a positive cosmological constant)
a much smaller primordial field would be sufficient for explaining the
present galactic fields.
The reason is simple:
The increased age of galaxies in such a universe gives the primordial
seed field more time to grow in strength.

In this article we shall concentrate on the case of the U(1) Higgs model,
partly to simplify the analysis but also to compare our results
with those of Kibble and Vilenkin \cite{KibVil}, who
developed a formalism for calculating the
magnetic fields generated in bubble collisions. Of course, one should not
really call it a magnetic field in the case of an Abelian gauge theory,
owing to the lack of a massless photon in the broken-symmetry phase.
Instead we will generally refer to it as a ``field strength''.
The authors of Ref.~\cite{KibVil} made
some potentially severe approximations in order to solve the
field equations: The modulus of the complex Higgs field was kept fixed in
the true vacuum, and the initial phase of the Higgs field at the moment of
collision was approximated by a step function. With these
basic
assumptions, they
were able to derive some fascinating analytical results concerning the
generation of the field strength in the collision region.
They demonstrated how the gauge-invariant current across the
interface of the two colliding bubbles produces
a ring of flux of the Abelian field strength
which encircles the bubble intersection region. On top of this they
defined a gauge-invariant phase difference between the two bubbles
and solved for its future evolution
to show that it
oscillates and eventually
relaxes to
zero as the bubbles merge. This example of
phase equilibration between two colliding bubbles
is in fact
a generic feature of models with either local or global symmetry
\cite{KibVil,Melfo}.

Here we go beyond the approximations introduced in
Ref.~\cite{KibVil}
and, by doing so, test their regime of validity. In particular, we allow for
a smoother interpolation
of the phase between the two bubbles,
using an approximation to the bounce solution for
a single bubble, and the fact that
we can represent the scalar field of two
well separated bubbles as a linear superposition
of the scalar fields for each bubble
\cite{Davis,SafCop}. Remarkably, we can obtain explicit
time-dependent expressions for both the
gauge-invariant phase and the Abelian
field strength. We compare these expressions
to those in Ref.~\cite{KibVil} as well as to the
numerical
solution of the field equations.
The results are very encouraging; we find incredible
agreement between our analytical solutions and the numerical solutions.
We also
demonstrate to what extent fields in
the collision region are
well
approximated by the step-function ansatz adopted in Ref.~\cite{KibVil}.

Real phase transitions naturally possess
lots of complicating factors which we do not take into account
here, such as frictional damping of the
bubble-wall velocity.
Particles that acquire a mass due to couplings to the Higgs field
are massless outside the bubble and massive inside. Not all the particles
external to the bubble have enough energy to become massive and must
reflect off the bubble wall, slowing it down \cite{Melfo,turok,lilley}.

Another complication is the conductivity associated with the plasma
surrounding the expanding bubble wall.
Charged particles in the plasma couple to the gauge fields and cause
them to lose their energy to the thermal bath.
This has the effect of reducing the likelihood of
forming loops of flux during the bubble collision as pointed out in
\cite{Davis,KibVil} and confirmed in \cite{CopSaf}. Furthermore,
the existence of a highly conductive plasma has the effect of conserving
magnetic flux by virtue of Faraday's and Ohm's laws.

In spite of these potential problems, we believe that the simple system
of colliding bubbles
is well worth investigating, not least because the field theory is highly
non-linear and yet
we are able to find extremely accurate analytical solutions.
Moreover, as we shall discuss in the conclusions, the method and results
can be generalised and applied to different areas, such as
the Standard Model and its extensions.

The paper is organised as follows. In Sec.~2 we review the standard
results due to Coleman \cite{Coleman} on bounce solutions and bubble
nucleation
in first-order phase transitions.
Following Ref.~\cite{Garriga}, we also investigate the evolution of bubbles
of arbitrary nucleation radius and find that all expanding bubbles
are well approximated by the bounce solution after the bubble size has
increased only by a small amount.
In Sec.~3 we review the key
approximations and results obtained in Ref.~\cite{KibVil}, in particular the
solutions for the gauge-invariant
phase and the field strength arising from the step-function
approximation. We rewrite these solutions, and solutions for other
gauge-invariant quantities, as closed-form expressions in
terms of Bessel functions.
In Sec.~4 we obtain improved solutions by using
the smooth bounce solution for the scalar field of each individual bubble
and the linear superposition ansatz for the combined
scalar field.
We derive explicit time-dependent solutions for
the gauge-invariant phase and the Abelian field strength.
The numerical lattice computation is described in Sec.~5.
In Sec.~6
we compare the numerical field evolution to our new smooth solution
and to the results obtained in
Ref.~\cite{KibVil}. Of particular
note is that the step-function approximation leads to
discontinuities in the field strength and in the phase
that are not there in the smooth
approximation. Moreover, the energy density associated with the gauge field
arising from the step-function approach is about
sixteen times larger than the true solution found
both numerically and in the smooth approximation.
We conclude in Sec.~7.

\section{Kinematics of Expanding Spherical Bubbles}

In a pioneering paper Coleman \cite{Coleman} investigated
the physics of bubble nucleation in the decay of the metastable vacuum.
He obtained the so-called bounce solution for the modulus of the scalar
field corresponding to a single nucleated bubble.
If the bubble radius is large compared to
the wall thickness, the bubble wall
can equivalently be described as
a relativistic membrane. The motion of such a membrane
subjected to the
outward
pressure caused by the energy difference of the
two vacua can be derived and was investigated in the spherical case
by Garriga \cite{Garriga}.
Below we will reproduce a number of key results in
Refs.~\cite{Coleman,Garriga},
extracting details of the bubble-wall dynamics that
we shall require later.
Of particular use will be an analytic approximation, Eq.~(\ref{rhoapprox1}),
to the actual
bounce solution, Eq.~(\ref{bubsol1}),  as well as
the most probable size of the nucleated
bubble, Eq.~(\ref{R0value}),
and the associated wall thickness, Eq.~(\ref{thick1}).

We shall see below that bubbles created via an instanton description of
tunnelling have O(1,3) symmetry. When there
are two bubbles present the spatial rotation symmetry is reduced to
O(2), so in fact the two-bubble collision has an O(1,2) symmetry
\cite{Hawking}. When we have two bubbles there is the chance that they
may nucleate at different times and so have different sizes when they
collide. We may however always boost to a frame where they nucleate
simultaneously; this is the frame in which we shall work.

\subsection{Bubble Nucleation}

\label{nuclsec}
For simplicity we choose a
Higgs potential\footnote{We remark that the choice of potential is not
important, and that the phenomena discussed in this article are qualitatively
similar for any effective potential exhibiting a first-order phase
transition. The quantitative results will be seen to
depend on generic features
of the potential, such as barrier height, the
location of the two minima, and their difference in energy density.}
that is cubic in $\phi^{\ast}\phi$ \cite{Hawking}:
\begin{equation}
\label{pot}
V(|\phi|)=\lam(|\phi|^2-\eta^2)^2\left(\frac{|\phi|^2}{\eta^2} + \eps
\right) - \eps\lam\eta^4~,
\end{equation}
where $\phi$ is the scalar Higgs field and
$0<\eps<\frac{1}{2}$. It has a metastable minimum at
$\phi=0$ and a degenerate global minimum at $\phi=\eta e^{i\theta}$,
$\theta\in [0,2\pi)$, separated by a barrier of height
$\left[4(1+\eps)^3/27-\eps\right]\lam\eta^4$. The difference in energy
density between the two minima is
$p=\eps\lam\eta^4$, and this constant has been subtracted in
Eq.~(\ref{pot}) in order to obtain zero energy density
in the metastable minimum.
The Higgs-boson mass $M_H$ in this model
is given by $M_H^2=4(1+\eps)\lam\eta^2$.

The nucleation of a bubble of true Higgs vacuum $|\phi|=\eta$
occurs by means of tunnelling
from $\phi=0$ through the barrier \cite{Coleman}.
The tunnelling is expected to be dominated by solutions to the
Euclidean equations of motion, with a nucleation probability
$\propto\exp(-\se)$,
where $\se$ is the Euclidean action. $\se$ is given by
\begin{equation}
\label{sedef}
\se[\phi,A_\mu]=\int d^4\bar{x}\left[
|D_\mu\phi|^2 + V(|\phi|) +
\frac{1}{4}F_{\mu\nu}F^{\mu\nu}
\right],
\end{equation}
where a positive definite metric is used.
The covariant derivative is  $D_\mu=\partial_\mu+ieA_\mu$ with
$F_{\mu\nu}=\partial_\mu A_\nu-\partial_\nu A_\mu$, and $A_\mu$ is the
associated U(1) vector potential.
Tunnelling is therefore dominated by those instantons with lowest
Euclidean action, something which is achieved by having vanishing
vector potential and a constant phase for the scalar field $\phi$.

\begin{figure}[htb!]
  \begin{center}
    \begin{minipage}{11cm}
       \begin{center}
       \leavevmode
       \epsfxsize=9cm \epsfbox{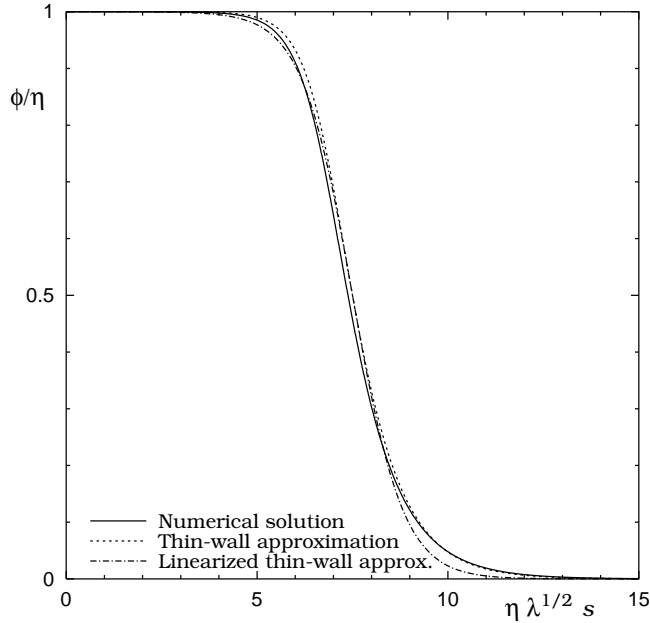}
       \end{center}
\caption{Bounce profiles for $\eps=0.2$. Here are shown
the exact solution of Eq.~(\ref{eucleq})
(solid),
the thin-wall approximation given by Eq.~(\ref{bubsol1})
(dotted), and the ``linearized'' thin-wall approximation given by
Eq.~(\ref{rhoapprox1}) (dot-dashed).}
\label{bouncefig}
    \end{minipage}
  \end{center}
\end{figure}

In the absence of gauge fields the instantons which have lowest action
(\ref{sedef}) are those with
O(4) symmetry, i.e. $\phi=\phi(s)$, where $s=\sqrt{\te^{\,2} +
\bm{x}^2}$ and $\te=it$ is the Euclidean time coordinate.
In fact, all other solutions to the Euclidean equations are divergent
\cite{gidas79}.
In considering the nucleation of one isolated bubble we may,
without loss of generality,
choose $\phi$ to be real and non-negative.
For such configurations the $\phi$ equation of motion is
\begin{equation}
\label{eucleq}
\frac{1}{s^3}\frac{d}{ds}\left( s^3 \frac{d\phi}{ds}\right)
 = \frac{1}{2}\frac{dV}{d\phi}~.
\end{equation}
The ``bounce'' solution \cite{Coleman} to this equation
satisfies the boundary conditions
 $\phi'(0)=0$, $\phi(\infty)=0$ and
is characterised by
a kink, the bubble wall, localised near $s=R_0$, where $R_0$ is
to be determined. For $\eps\ll 1$, $R_0$ turns out to be
large compared to the width of the kink,
so one can set $s\approx R_0$ in the
region where $d\phi/ds$ differs
significantly from zero.
Moreover, for small $\eps$
we have $V(\phi)\approx V_0(\phi)$, where
$V_0(\phi)$ is obtained from $V(\phi)$ by setting $\eps=0$, and
$\phi(0)\approx\eta$. With these approximations,
collectively
known as the ``thin-wall'' approximation, Eq.~(\ref{eucleq})
is easily integrated to yield
\begin{equation}
\label{fstint1}
\frac{d\phi}{ds} = -\sqrt{V_0(\phi)}~.
\end{equation}
After one further integration one obtains the bounce
solution particular to our potential,
\begin{equation}
\label{bubsol1}
\phi_{\rm b}(s) =\frac{\eta}{\sqrt{2}}\left[
1 - \tanh\left(\eta \sqrt{\lam}(s-s_0)\right)
\right]^{\frac{1}{2}}~,
\end{equation}
where a constant of integration $s_0$ appears because
the boundary condition
$\phi'(0)=0$ is not exactly satisfied in the thin-wall
approximation.
The bounce profile
is centred at the value of $s$ for which $\phi_{\rm b}=\eta/2$,
i.e.\  $s=R_0\equiv s_0 +
\arccoth(2)/(\eta\sqrt{\lam})$.
Near $s=R_0$ the bounce solution is well approximated by
\begin{equation}
\label{rhoapprox1}
\phi_{\rm b}(s)\simeq \frac{\eta}{2} [1 - \tanh 2\mu_0(s-R_0)]~,
\end{equation}
where
\begin{equation}
\label{mu0def}
\mu_0\equiv -\frac{1}{\eta}\phi_{\rm b}{}'(R_0)=\frac{1}{\eta}
\sqrt{V_0(\eta/2)}=\frac{3 \eta\sqrt{\lam}}{8}\approx
\frac{3M_H}{16}
\end{equation}
by virtue of Eq.~(\ref{fstint1}). Asymptotically, however,
the approximation (\ref{rhoapprox1})
differs
from
the solution (\ref{bubsol1}).
Whereas the latter tends to zero
as  $\phi_{\rm b}\sim
\exp(-\eta\sqrt{\lam}s)$ for $s\to +\infty$ and tends to $\eta$
as $(\eta-\phi_{\rm b})\sim
\exp(2\eta\sqrt{\lam}s)$ for $s\to -\infty$, the approximation
(\ref{rhoapprox1}) instead approaches both these asymptotic values as
$\sim
\exp(-3\eta\sqrt{\lam}|s|/2)$.
Fig.~\ref{bouncefig} shows a comparison of
the analytical approximations
of the bounce profile
given in Eqs.~(\ref{bubsol1}) and (\ref{rhoapprox1})
with
the numerical solution of Eq.~(\ref{eucleq}).

The nucleation radius $R_0$ is as yet undetermined, but can be
calculated by extremising the Euclidean action.
For the bounce solution in the absence of
gauge fields this action can be calculated using
the thin-wall approximation and Eq.~(\ref{fstint1})
without reference to the solution itself.
One obtains
\begin{equation}
\label{sesol}
\se\simeq 2\pi^2\left(\sigma R_0^3 - \frac{1}{4} p R_0^4\right)~,
\end{equation}
where
\begin{equation}
\label{sigmadef}
\sigma = \int_0^{\infty} ds\left[\left(\frac{d\phi}{ds}\right)^2
+V(\phi)\right]
\simeq 2\int_0^\eta d\phi \sqrt{V_0(\phi)}
=
\eta^3 \sqrt{\lam}/2~,\quad\quad p=\lam\eps\eta^4~.
\end{equation}
Here $\sigma$
is the surface tension, or mass per unit area, of the bubble wall,
whereas the outward pressure $p$ is equal to
the difference in
energy density between the metastable and global minima.
The factor $2\pi^2R_0^3$ in Eq.~(\ref{sesol})
is the area of the three-sphere, whilst
$2\pi^2 R_0^4/4$
is the four-volume of the instanton.
The action $\se$ is extremised by the value,
\begin{equation}
\label{R0value}
R_0=\frac{3\sigma}{p}= \frac{3}{2\eps\eta\sqrt{\lam}}~.
\end{equation}

The initial conditions for the evolution of a bubble is given
by the $t_E=0$ slice of the instanton, along which
\mbox{$\partial\phi_b/\partial t_E=0$}, such that the bubble
is initially stationary. The future evolution is found by analytically
continuing the bounce solution to Minkowski space; details may be found
in Ref.~\cite{Coleman}.

The thin-wall limit allows us to make statements about the
evolution as well as the nucleation of bubbles. A
measure of the bubble-wall thickness is provided by
$\mu^{-1}$, where
\begin{equation}
\label{thick1}
\mu\equiv-\frac{1}{\eta}\left.\frac{d\phi_{\rm b}}{dr}\right|_{s=R_0}=
\mu_0\frac{r(t)}{R_0} = \frac{\mu_0}{\sqrt{1-\dot{r}^2}}~,\quad\quad
\dot{r}\equiv\frac{dr(t)}{dt}
\end{equation}
showing that the thickness of the
bubble wall is subject to the expected Lorentz
contraction.
The thickness $\mu_0^{\,-1}$ at $t=0$ is given
in the thin-wall approximation by Eq.~(\ref{mu0def}).
By means of Eqs.~(\ref{mu0def})
and (\ref{R0value}) the condition for the thin-wall approximation
to be valid can
be stated
\begin{equation}
\label{thincond}
\mu_0 R_0 = \frac{9}{16\eps} \gg 1~.
\end{equation}

\subsection{Spherical Membrane Dynamics}
\label{sphermem}
Bubbles whose time evolution is not O(1,3) symmetric are nucleated,
but with lower probability,
as the result of tunnelling via instantons that do not extremise
the Euclidean action.
In particular, such bubbles may deviate from spatial spherical
symmetry, or they may be spherically symmetric but not invariant under
boosts.  We shall here consider the latter possibility.

When
the bubble wall is thin compared to
the radius of the bubble, it can be approximated by an infinitely thin
membrane. The equation for the radius $r(t)$
of a spherical membrane subject to an outward pressure can be
written \cite{Garriga}
\begin{equation}
\label{sphmemb}
\frac{dE}{dt}=0~,\quad
E\equiv\frac{\sigma}{\sqrt{1 - \dot{r}^2}}r^2 - \frac{1}{3}
p r^3~,
\end{equation}
where $\dot{r}=dr/dt$, $\sigma$ is the surface tension, or mass per unit
area, of the
membrane and $p$ is the difference in energy density between the false
vacuum exterior to the bubble and the true vacuum inside. Equation
(\ref{sphmemb}) simply expresses the conservation of energy, but
can also be obtained as a special case of the motion of
a relativistic membrane in an external force field.
Because of the relativistic factor $(1-\dot{r}^2)^{-1/2}$ the first term
of $E$ comprises the total energy of the bubble wall, mass as
well as kinetic energy, while the second term is the total energy of the
true vacuum inside the bubble.
The general solution $r(t)$ is given implicitly as
\begin{equation}
\label{rimplic}
\int_{r(0)}^{r(t)} dr \frac{3 E + p r^3}{\sqrt{
(3 E + p r^3)^2 - 9 \sigma^2 r^4}} = \pm t~.
\end{equation}
In the special case $E=0$, one obtains an explicit solution with O(1,3)
symmetry,
\begin{equation}
\label{o31sol}
r^2 - t^2 = \ri^2~,\quad \ri=r(0)~.
\end{equation}
If the bubble is nucleated at rest at $t=0$, we have from
Eq.~(\ref{sphmemb}) that its initial radius is $\ri=R_0\equiv
3\sigma/p$. Equation (\ref{o31sol}) then coincides with the equation
of motion for the centre of the bubble wall in the O(1,3)
Wick-rotated bounce solution considered above.

\begin{figure}[htb!]
  \begin{center}
    \begin{minipage}{11cm}
       \begin{center}
       \leavevmode
       \epsfxsize=9cm \epsfbox{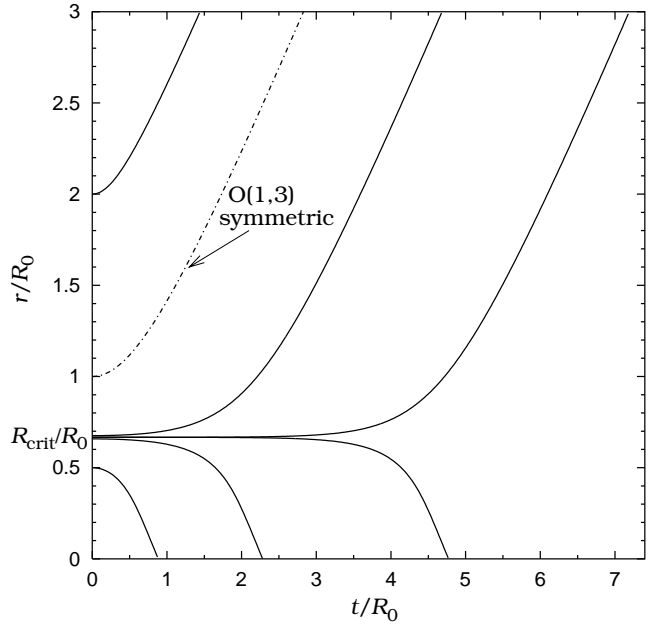}
       \end{center}
\caption{Evolution of the radius
of spherical bubbles nucleated at rest in a first-order phase
transition. $R_0=3\sigma/p$.}
\label{trajfig}
    \end{minipage}
  \end{center}
\end{figure}

A set of bubble-wall trajectories for different values
of $E$ are depicted in Fig.~\ref{trajfig}. The bubbles are all chosen to
start
from rest at $t=0$. If the initial radius $\ri$ is smaller than
the critical radius
$R_{\rm crit}=2\sigma/p=(2/3)R_0$, the bubble recollapses, whereas
if $\ri>R_{\rm crit}$, it expands. Only the trajectory with $\ri=R_0$
is a hyperbola, corresponding to the O(1,3) symmetry. It should
be noted that, although the
initial expansion for $\ri>R_{\rm crit}$ takes place on
different timescales, once the bubble radii have increased by about
$R_0/2$ the remaining evolution is quite well approximated by
a time-translated O(1,3) solution. Therefore, if the average distance
between nucleation sites is larger than about $3 R_0$, the
O(1,3)-symmetric solution will generally
provide a good description of
expanding bubbles at the time of their collision.

\section{Bubble Collisions and the Kibble-Vilenkin Approximation}
\label{bubcol}

We now proceed to study collisions of two expanding bubbles
of true vacuum. As was argued in Sec.~\ref{nuclsec}, for typical
bubbles the phase of the
complex field inside each bubble is (covariantly) constant, and
we may initially choose a gauge in which the gauge vector potential
vanishes everywhere. Then there will in general be a random initial
phase difference $\Delta\theta_0$ between the two bubbles.

It was shown by Kibble and Vilenkin \cite{KibVil} that, because of
the phase
difference, a gauge-invariant current
$j_k=i e [\phi^{\ast}D_k \phi - (D_k\phi)^{\ast}\phi]$ develops across
the surface of intersection of the two bubbles as they collide.
The current gives rise to a ring-like flux of the Abelian field strength
$F_{ij}$ which takes the shape of a girdle encircling the bubble
intersection region. Moreover, these authors showed that,
under the assumptions that the radial mode of the scalar field is
strongly damped and a step function can be used
to approximate the initial phase
of the complex field at the moment of collision, the phase
oscillates and equilibrates to a uniform value as the bubbles merge.

We shall here review briefly the approach of Ref.~\cite{KibVil},
as it forms a common basis for the analytical methods discussed and
developed in this article. We will analyse the results
arising from their approximation and derive closed-form
expressions for
gauge-invariant quantities such as the Abelian
field strength. Crude estimates are extracted for
the peak value of the field strength and the scale of spatial variation
of the field. These quantities are relevant for determining the
strength and correlation length of magnetic fields that may be
created through bubble collisions in the early universe
in more realistic models.

\begin{figure}[htb!]
  \begin{center}
    \begin{minipage}{11cm}
       \begin{center}
       \leavevmode
       \epsfysize=11.2cm \epsfbox[200 20 285 435]{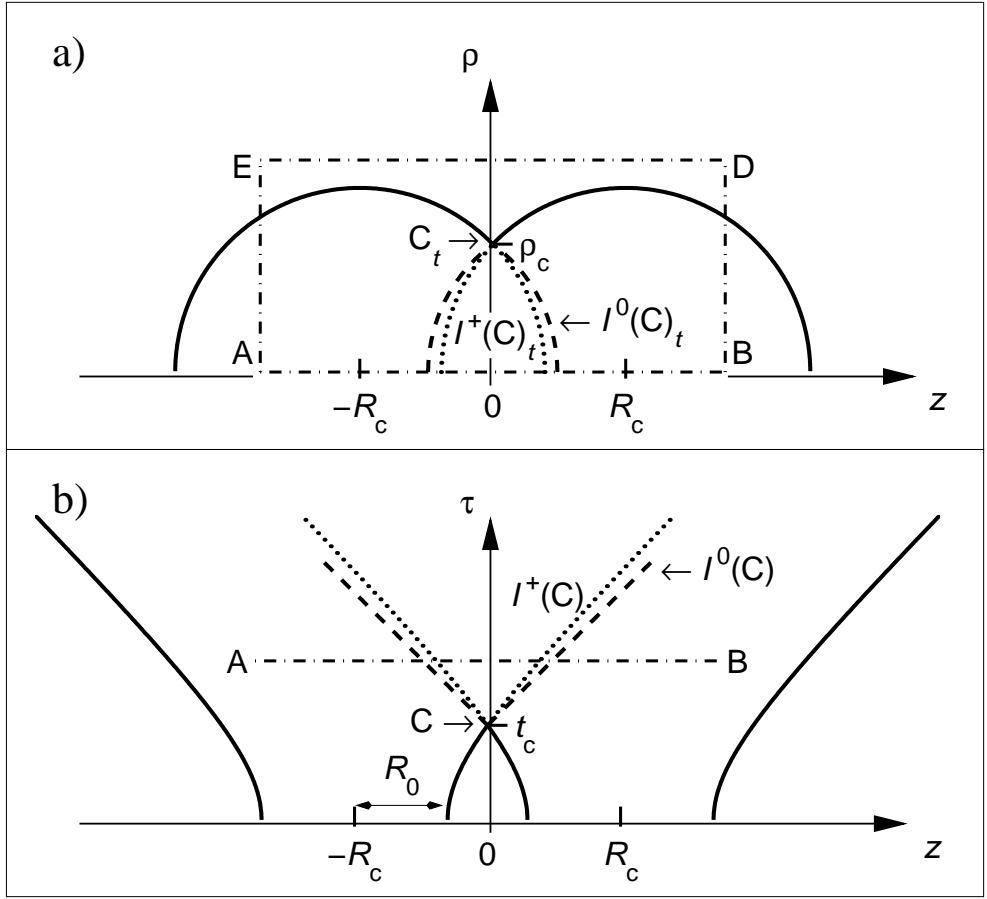}
       \end{center}
\caption{Bubble collision in two different coordinate systems.
Dotted curves indicate the extrapolated position of each bubble wall
in the absence of a collision.
The union of causal futures of bubble-wall intersections, $I^{+}({\rm C})$,
extends out to the causal horizon
$I^{0}({\rm C})$ drawn as a dashed curve, and is the only region where
physical fields that result from bubble
collision events can exist. Further notation is
explained in the text. (a)
Slice of bubble collision for constant time $t$ and constant
azimuthal variable $\varphi$. Here ${\rm C}_t$ denotes the
instantaneous
position of the circle of most recent intersection, whose radius is
$\rho_{\rm c}$. The contour ABDEA is used for evaluating the Abelian flux.
(b) Reduced spacetime diagram of entire bubble collision. The point C
represents all intersections of the two bubble walls.}
\label{drawing}
    \end{minipage}
  \end{center}
\end{figure}

Let us consider two non-overlapping bubbles, each locally O(1,3)
symmetric with respect to a frame with origin at its centre.
The separation of their nucleation events is then necessarily spacelike,
and one may pick a reference frame and coordinate origin such that
they are nucleated simultaneously with centres
at $(t,x,y,z)=(0,0,0,\pm \rc)$. In this
frame, the bubbles have equal initial radius $\ri=R_0$.
Their first collision occurs at
$(\tc,0,0,0)$ when their radii are $\rc$ and
$\tc=\sqrt{\rc^2 - R_0^2}$ by virtue
of Eq.~(\ref{o31sol}).
\footnote{Unlike Ref.~\cite{KibVil} we
do not assume that $\tc/\rc=1$,
which would be approximately true only if $\rc\gg R_0$.}
The system of two bubbles has the invariance
O(1,2) reducing the problem to a $1+1$ dimensional one
that can be expressed in terms of the coordinates $z$ and $\tau$,
where
\begin{equation}
\label{taudef}
\tau^2 = t^2-x^2-y^2~.
\end{equation}
In these coordinates the point of first contact $z=0$, $\tau=\tc$
 (denoted by C in Fig.~\ref{drawing}b) represents the worldsheet of
the circle of most recent intersection
of the bubble walls in $3+1$ dimensional spacetime. This circle
recedes superluminarly from the $z$ axis according to the equation
$x^2+y^2=\rho_{\rm c}^{\,2}\equiv t^2-\tc^2$, $t\geq \tc$. Its
 position at a particular time $t$ is indicated by the label
${\rm C}_t$ in Fig.~\ref{drawing}a.

 Let us write the scalar Higgs field in polar form $\phi=Xe^{i\theta}$.
In the axial gauge $A_z=0$ its modulus, phase, and
the vector potential take the form
\begin{equation}
\label{thetaa}
X=X(\tau,z),\quad
\theta(x)=\tha(\tau,z), \quad A^{\alpha}(x)=x^\alpha a(\tau,z)~,
\end{equation}
where $\alpha\in \{0,1,2\}$. The condition $\tha(\tau,0)=0$
fixes the gauge completely.

Following Ref.~\cite{KibVil} the modulus $X$ of the scalar
field inside the bubbles and in their overlap region is approximated
by the constant true-vacuum value $\eta$.
The equations for
$\tha(\tau,z)$ and $a(\tau,z)$ then reduce to the linear equations
\begin{eqnarray}
\label{thetalineq}
\partial_\tau^2\tha + \frac{2}{\tau} \partial_\tau\tha -
\partial_z^2\tha +m^2\tha &=& 0~,\\*
\label{alineq}
\partial_\tau^2 a + \frac{4}{\tau} \partial_\tau a -
\partial_z^2 a +m^2 a &=& 0~,
\end{eqnarray}
together with
\begin{equation}
\label{fstordereq}
3 a + \tau\partial_\tau a= \frac{m^2}{e} \tha~,
\end{equation}
where $m$ is the vector-boson mass, $m^2=2e^2\eta^2$.
The constant-modulus approximation is not quite consistent, as
nontrivial solutions of Eqs.~(\ref{thetalineq}) and
(\ref{alineq}) do not satisfy the remaining
equation
\begin{equation}
\label{Xeq}
\Box X - (D_\mu\theta D^\mu\theta) X + \frac{1}{2}\frac{dV}{dX} = 0
\end{equation}
for constant $X\equiv\eta$. On the contrary,
non-trivial gauge-invariant phase gradients $D_\mu\theta\equiv
\partial_\mu\theta + e A_\mu$ appear inside the
causal future of bubble-wall intersections
\mbox{$I^{+}({\rm C})$} $=$ \mbox{$\{(z,\tau)\,|\,
\tau-\tc>|z|\}$} (see Fig.~\ref{drawing}b) and serve
as sources for fluctuations in the Higgs-field modulus. Fortunately,
for a small initial phase difference
$\Delta\theta_0$ the fluctuations in $X$ are also small, so that
Eqs.~(\ref{thetalineq}), (\ref{alineq}) and (\ref{fstordereq}) remain
a good approximation for the phase and
the vector potential.

In this article we shall be interested mainly in two quantities,
the Abelian field strength
$F^{\alpha z}=x^{\alpha}\partial_z
a(\tau,z)$ and the phase $\theta$ of the complex field.
The field strength is gauge invariant,
but the phase is not. Because the analytical treatment simplifies
in the axial gauge, while the lattice computations of Sec.~\ref{latsec}
are conveniently performed in the temporal gauge, it is useful to
construct
a gauge-invariant generalisation of the phase $\theta$ which can be
calculated in either gauge, so that the independent
results may be compared. Such a gauge-invariant phase can be defined
\cite{KibVil} at any point P by means of the line integral of the
gauge-invariant phase gradient
$D_\mu\theta\equiv \partial_\mu\theta+eA_\mu$ from
the origin to P. Unfortunately,
this line integral is path dependent.
In this work we shall fix the definition
of the gauge-invariant phase $\tilde{\theta}$
by specifying a path from the origin as follows:
\begin{equation}
\tilde{\theta}(\tau,z) \equiv \int_{(0,0,0,0)}^{(t,x,y,0)} dx^\alpha D_\alpha
\theta
+  \int_{(t,x,y,0)}^{(t,x,y,z)} dz D_z\theta ~,
\label{giphase}
\end{equation}
where $\alpha=0,1,2$ is summed over.
By choosing initial conditions
for Eqs.~(\ref{thetalineq}) and (\ref{alineq})
(or the full, nonlinear equations)
that adhere to the symmetry of the problem, one
can arrange for the solution to be such at all times that
$\partial_\alpha\theta$ and $A_\alpha$  are both
odd functions of the variable $z$. Then the first term of Eq.~(\ref{giphase})
does not contribute, and we have
\begin{equation}
\tilde{\theta}(\tau,z) \equiv
\int_{(\tau,0)}^{(\tau,z)} dz D_z\theta(\tau,z) ~.
\label{giphase2}
\end{equation}
The expression simplifies further in the axial gauge $A_z=0$, where one
finds $\tilde{\theta}(\tau,z)\equiv\tha(\tau,z)$.
Note that, while $D_\mu\theta$ is defined locally,
the phase $\tilde{\theta}$ is a non-local quantity by construction.
Therefore, it is not a physical variable, and the evolution of
$\tilde{\theta}$ exhibits non-causal correlations. Nevertheless,
$\tilde{\theta}$ provides a useful, gauge-invariant characterisation
of the field configuration.

\subsection{Step-Function Approximation}
\label{step}

In the approach by Kibble and Vilenkin \cite{KibVil}, a sign function
is used\footnote{A similar approach is adopted in Ref.~\cite{Davis}.}
 to approximate the initial phase difference between the two
bubbles at the instant when they collide, $\tau=\tc$. This would
correspond to bubbles with infinitely thin walls. The
initial conditions on the phase and its derivative are taken to be
$\tha(\tc,z)= \theta_0\,\sgn(z)$,
corresponding to a phase difference of $\Delta\theta_0=2\theta_0$,
and $\partial_\tau\tha(\tc,z)=0$. The vanishing of the derivative
can be understood from Fig.~\ref{drawing}b, since at
$\tau=\tc$ points with $|z|>0$ are in the interior of bubbles with
constant phase $\pm\theta_0$, and $\tha(\tau,0)=0$ from the gauge condition.
In this ansatz, the bubbles are treated as though they extend to
$z=\pm\infty$, whereas the solutions should strictly speaking
only be trusted
where the constant modulus approximation is good, i.e.\ in
the interior of the bubbles given by
$|z|<\rc +\sqrt{R_0^2 +\tau^2}$.
In practise, this does not matter, since gauge-invariant physical fields
only propagate inside the
causal future of bubble-wall intersections $I^{+}({\rm C})$.
This region extends
beyond the overlap region found by extrapolating the original bubble
trajectories (dotted lines in Fig.~\ref{drawing}),
but never exits the union of the interior of the two bubbles.

The vector potential is chosen to be zero initially, $a(\tc,z)=0$,
and its derivative
is determined from Eq.~(\ref{fstordereq}). One obtains
$\partial_\tau a(\tc,z)=m^2 \theta_0\,\sgn(z)/(e \tc)$.
The solution of the linearised equations (\ref{thetalineq}) and
(\ref{alineq}) for $\tau>\tc$
then
becomes \cite{KibVil}
\begin{eqnarray}
\label{Kibbletheta}
\tha(\tau,z) &=& \frac{\theta_0 \tc}{\pi\tau}\int_{-\infty}^\infty
\frac{dk}{k}\sin k z\left(\cos\omega(\tau-\tc) + \frac{1}{\omega\tc}
\sin\omega(\tau-\tc)\right)~,\\*
\label{Kibblea}
a(\tau,z)&=& \frac{\theta_0 m^2\tc}{\pi e\tau^3}
\int_{-\infty}^\infty
\frac{dk}{k}\sin k z\left[
-\frac{\tau-\tc}{\omega^2\tc}\cos\omega(\tau-\tc)\right.\nonumber\\*
&&\hspace*{5cm}\left. +
\left(\frac{\tau}{\omega}+\frac{1}{\omega^3\tc}\right)
\sin\omega(\tau-\tc)\right]~,
\end{eqnarray}
where $\omega^2=k^2+m^2$.
These integrals are highly oscillatory, but can be put in a form that
greatly facilitates their numerical evaluation. Letting $T=\tau-\tc$
we have
\begin{equation}
\label{thetanice}
\tha(\tau,z)= \left\{
\begin{array}{ll}
\displaystyle\theta_0 \frac{\tc}{\tau} \int_0^z du\left(
\frac{1}{\tc}J_0(m\sqrt{T^2-u^2}) - mT
\frac{J_1(m\sqrt{T^2-u^2})}{\sqrt{T^2-u^2}}
\right) &,~ |z|< \tau-\tc \\*[3ex] \displaystyle
\theta_0 \frac{\tc}{\tau}\,\sgn(z)\left(\cos mT +\frac{1}{m\tc}\sin mT\right)
&,~ |z|>\tau-\tc~. \end{array}\right.
\end{equation}
Similarly for the vector potential $A^\alpha=x^\alpha a$ we obtain
\begin{equation}
\label{aoutside}
a(\tau,z)=\theta_0 \frac{m^2}{e}\frac{\tc}{\tau^3}\,\sgn(z)\left[
-\frac{T}{m^2 \tc}\cos mT + \left(\frac{\tau}{m} + \frac{1}{m^3\tc}\right)
\sin mT\right]~,~~ |z|>\tau-\tc
\end{equation}
or, for general $z$,
\begin{equation}
\label{atrivial}
a(\tau,z)= \frac{1}{\rho}\int_0^z du B^\varphi(\tau,u)~,
\end{equation}
where $\rho=\sqrt{x^2+y^2}$ is the distance from the $z$ axis, and
the azimuthal Abelian field strength $B^\varphi=F^{z\rho}=\rho\partial_z a$
has the closed-form expression
\begin{equation}
\label{Bclosed}
B^\varphi(\tau,z)= \left\{
\begin{array}{ll}
\displaystyle \rho \frac{\theta_0 m^2\tc}{e\tau^3}\left(
\tau J_0(m\sqrt{T^2-z^2}) +\frac{\sqrt{T^2-z^2}}{m\tc}
J_1(m\sqrt{T^2-z^2})\right) &,~ |z|< \tau-\tc \\*[3ex]
\displaystyle
0
&,~ |z|>\tau-\tc~. \end{array}\right.
\end{equation}
The corresponding ``electric field'' $E^z$ is axial and is given by
$E^z=F^{0z}=
-t\partial_z a = -(t/\rho) B^\varphi(\tau,z)$.
The gauge-invariant phase gradient $D_\mu\theta$ can also be written
in closed form; in particular $D_z\theta=\partial_z\tha$ is obtained
directly from Eq.~(\ref{thetanice}). For $\alpha\in\{0,1,2\}$ one finds,
using Eqs.~(\ref{alineq})
and (\ref{fstordereq}),
\begin{equation}
\label{covalpha}
D_\alpha\theta  = \left\{
\begin{array}{ll}
\displaystyle x_\alpha z \frac{\theta_0 \tc}{\tau^3}\left(
m\tau \frac{J_1(m\sqrt{T^2-z^2})}{\sqrt{T^2-z^2}}-
\frac{1}{\tc}J_0(m\sqrt{T^2-z^2})
\right) &,~ |z|< \tau-\tc \\*[3ex]
0&,~ |z|>\tau-\tc~. \end{array}\right.
\end{equation}

The physical fields
$B^\varphi$, $E^z$, and
$D_\mu\theta$
are nonzero only in the causal
future of bubble-wall intersections $I^{+}({\rm C})=\{(z,\tau)\,
\Big|\,
|z|<\tau-\tc\}$.
In the 3+1 dimensional coordinates at a particular time $t$
this corresponds to a region
enclosed by
two rotational paraboloids shown as dashed curves
in Fig.~\ref{drawing}a. The physical fields are generated by
sources on the superluminarly
expanding circle ${\rm C}_t$ of most recent intersection
($\rho=\rho_{\rm c}\equiv \sqrt{t^2 -\tc^2}$, $z=0$) and
propagate towards the $z$ axis, i.e. {\em into\/} the bubble overlap
region.

One characteristic of the step-function approximation is that the original
discontinuity of $2\theta_0$ in the initial condition for the phase
splits after the collision into two equal discontinuities
of magnitude $\theta_0\tc/\tau$
at $z=\pm (\tau-\tc)$.
Similarly, the
Abelian field strength has a discontinuity on
$I^{0}({\rm C})$;
the limit from the interior is
$B^\varphi_{|z|\to (\tau-\tc)_{-}}=\theta_0 m^2
\tc \rho/(e \tau^2)$, while it is zero outside $I^{+}({\rm C})$.

On the plane $z=0$ the field
strength $B^{\varphi}$ oscillates about zero with an amplitude
which is an increasing [decreasing] function of $\rho$ [$\tau$].
The largest field strength anywhere
 is found on the circle ${\rm C}_t$ of most recent
intersection
($\rho=\rho_{\rm c}$, $z=0$) and has the value
 $|B_{\rm max}|=\theta_0
m^2 \rho_{\rm c}/(e\tc)$. As this circle expands,
$|B_{\rm max}|\propto t
\approx r(t)$ at late times $t$, where $r(t)$ is the
bubble radius.
The field strength $B^\varphi$
around the maximum defines a peak,
the radial width of which,
$\Delta\rho$, can be estimated from the location of the first zero
of $B^\varphi(\tau,0)$ for $\tau>\tc$ (i.e.\ $\rho<\rho_{\rm c})$.
With $m(\tau-\tc)=x_0$ at this zero one finds
\begin{equation}
\label{deltarho}
\Delta\rho=
\rho_{\rm c}
-\sqrt{t^2 - (\tc+\frac{x_0}{m})^2} \sim
\frac{x_0 (2 m \tc + x_0)}{2 m^2 t}~,
\end{equation}
where $x_{01}<x_0<x_{11}$ and $x_{n1}$ is the smallest positive zero
of the Bessel function $J_n(x)$, $x_{01}\approx 2.405$,
$x_{11}\approx 3.832$. The last expression in Eq.~(\ref{deltarho})
is the leading behaviour at large times $t$. The narrowing
of the peak with time is a relativistic effect.
For a fixed value of $\tau$, the field-strength peak
extends in the $z$ direction across
$I^{+}({\rm C})$,
so that the width in that direction can be
estimated as $\Delta z= x_0/m$.

We can now derive a crude estimate of the Abelian flux
contained in the maximal peak by taking
\begin{equation}
\label{crudeflux}
\Phi_{B,{\rm peak}}\approx |B_{\rm max}|\,\Delta \rho\,\Delta z
\sim \frac{\theta_0}{e} x_0^{\,2}\left(1+\frac{x_0}{2 m \tc}\right)~,
\end{equation}
where the last expression is the constant flux
approached at late times.
One can compare this to the total flux of $B^\varphi$ through
a semi-infinite plane based on the $z$ axis (i.e.\ the total flux of
all field lines that
``run around the $z$ axis''). This flux, $\Phi_{B}$,
is obtained
 by evaluating the line integral of $A_\mu$ at constant
time $t>\tc$
along the path ABDEA indicated in Fig.~\ref{drawing}a. In the
axial gauge $A_z=0$, the segments AB and DE do not contribute to the
integral.
In order to enclose all flux of the Abelian field strength, the segments
BD and EA must be chosen so that
their $z$ coordinates satisfy $|z|>t-\tc$,
and the points D and E must be located at
$\rho>\rho_{\rm c}$ (i.e.\ $\tau<\tc$).
Assuming that the vector potential $A_k$ vanishes for $\tau<\tc$,
in accordance with the initial condition, only $\rho$ such that
$0<\rho<\rho_c$ contribute to the integral. From Eq.~(\ref{aoutside})
we find
\begin{eqnarray}
\label{totflux}
\Phi_B(t)&=&-\left(\int_{\rm BD}+\int_{\rm EA}\right)dx^\alpha A_\alpha=
 2 \int_{\tc}^t d\tau\,\tau a(\tau,|z|)\nonumber\\*
&=&\frac{2\theta_0}{e}\left[
1-\frac{\tc}{t}\left(\cos m(t-\tc)) +\frac{1}{m\tc}\sin m(t-\tc)
\right)
\right]~, \quad t>\tc~.
\end{eqnarray}
Note that the integral receives no contributions from regions where
the Higgs-field modulus $X$ is fluctuating, and therefore the expression
(\ref{totflux}) for the total flux is valid also for large initial
phase difference $\Delta\theta_0$ of the bubbles (see discussion
after Eq.~(\ref{Xeq})). In this sense, the
total flux is of a topological nature.

As a function of time, the Abelian flux $\Phi_B$ rises from zero
and begins to perform damped oscillations about the late-time limit
$\Delta\theta_0/e$, where $\Delta\theta_0=2\theta_0$ is the initial
phase difference of the two bubbles. The amplitude of these oscillations
decreases as $t^{-1}$. We emphasise that the flux does
not go away; it is there to stay.

Comparing with Eq.~(\ref{crudeflux}) we see that the flux
$\Phi_{B,{\rm peak}}$ contained in
the maximal peak is
of the same order of magnitude as (or somewhat larger than\footnote{This
is not a contradiction, because $B^\varphi$ changes sign in the bubble
overlap region}) the
total flux $\Phi_B$. This shows that a significant
portion of the total flux is contained in an outward travelling
circular flux tube near $\rho=\rho_{\rm c}$, $z=0$ whose field strength
$B_{\rm max}$ is increasing with time.

\section{A New, Smooth Analytical Solution}

The step-function approximation
described in the previous
section
gives a rough picture of the
qualitative behaviour of
the physical fields in a U(1) bubble collision.
For instance, by looking ahead to Fig.~\ref{bsurf} one can see that
the shape of the
profile of the Abelian field strength in the bubble overlap region
is quite well reproduced.

On the other hand, the step-function approximation
predicts a peak value of the field strength that differs
by a factor of four from the correct value obtained
in numerical simulations (a factor of sixteen
difference in energy density). Moreover,
the solutions (\ref{thetanice})--(\ref{covalpha})
have discontinuities on the future null surface of bubble
intersections $I^{0}({\rm C})$ and are thus unable to account for the
behaviour of the fields near that surface.
Finally, the solutions are limited to
the region $\tau\geq\tc$ (where $\tau=\sqrt{t^2-\rho^2}$)
and therefore provide no clue to the field
behaviour shortly before the bubble collisions or, at later times,
outside the bubble walls.

Realistic bubble walls of finite width, such as given by
Eq.~(\ref{bubsol1}), have exponential tails that begin to overlap
long before the instant of collision. The overlapping tails initiate
the evolution of the phase of the scalar field and of the vector
potential so that, in contrast with the assumption made in
Sec.~\ref{step}, the functions $\partial_\tau\tha$ and
$A^\alpha=x^\alpha a$ are already
nonzero when the collision reaches full impact at $\tau=\tc$.

In this section, we shall model the initial evolution of $\tha$ and
$a$ and obtain a smooth analytical solution. To do so, we first
observe that the scalar Higgs field of two well-separated
bubbles can be written \cite{Davis,SafCop}
\begin{equation}
\label{linsup}
\Phi(x)=\Phi_1(x)+\Phi_2(x)~,
\end{equation}
where $\Phi_1$, $\Phi_2$ are O(1,3)-symmetric solutions of the Higgs
field equation with $\Phi_i\to 0$ away from the bubbles and $|\Phi_i|$
approaching the vacuum expectation value $\eta$ well inside each bubble.
In the absence of gauge fields the linear superposition (\ref{linsup})
remains a good approximation even when the bubbles begin to overlap,
because in the region of overlap the Higgs field is small and the
field equation can
be linearised about $\Phi=0$. When gauge fields are included, the
reliability of the linear superposition approximation for describing
the phase
of the Higgs field depends
on the choice of gauge. We have found that in the
{\em temporal gauge\/} the initial evolution of
the phase
is well described by a linear superposition approximation,
as long as the bubble overlap is weak.

The position of a bubble wall can be defined as the set of points where
$|\Phi_i|=\eta/2,~i=1,2$.
Note that this definition gives the location of each bubble wall in the
absence of the other bubble. The bubble collision is then defined as
an event $x$ for which $|\Phi_1(x)|=|\Phi_2(x)|=\eta/2$. This
corresponds to the intersection of the extrapolated trajectories of
two well-separated bubble walls. When the bubbles begin to
 overlap strongly,
the actual scalar field $\Phi(x)$ has a nonlinear behaviour
and therefore cannot be used to define the event of collision.

Using the same geometry as in Sec.~\ref{bubcol},
depicted in Fig.~\ref{drawing},
the assumptions underlying the present analytical treatment can be
formulated as follows: The bubble collision occurs at $\tau=\tc$,
$z=0$
with $\tau=\sqrt{t^2-\rho^2}$.
For $\tau\leq \tc$ the field evolution is
given by the linear superposition (\ref{linsup}). This defines the
fields
outside the bubble walls as well as before the time of
collision. For $\tau>\tc$
the scalar field is assumed to have constant modulus $|\Phi|=\eta$.
This reduces the problem to one of finding the solution
of the linearised
equations (\ref{thetalineq}) and
(\ref{alineq}) with initial conditions at $\tau=\tc$
provided by
the linear superposition.

Note that although linear superposition is an excellent approximation
for $\tau\ll\tc$, and the constant-modulus approximation
works very well for $\tau\gg\tc$, both approximations break down
near $\tau=\tc$. The approach taken here is therefore
one of joining two asymptotic solutions across a non-linear region
where no solutions are known.

Some further approximations are necessary. We assume that the
thin-wall approximation (\ref{thincond}) is valid. Moreover, we shall
require
that the
two bubbles overlap negligibly at the time of their nucleation, $t=0$.
At this time, the bubble walls are a distance $2(\rc-R_0)$ apart, and
the thickness of the bubble wall is given by $\mu_0^{-1}$ (refer to
Eqs.~(\ref{mu0def}) and (\ref{thick1})). One obtains the
{\em no-overlap condition\/}
\begin{equation}
\label{nooverlap}
2 \mu_0(\rc-R_0)\gg 1~.
\end{equation}
Because of the exponential decay of the bubble-wall profile,
results accurate to about 3\% are obtained already when the left-hand
side of Eq.~(\ref{nooverlap}) is greater than 3.5. Via
the identity $\tc^2=\rc^2-R_0^2$ the condition of
no overlap also implies the following very useful relation,
\begin{equation}
\label{betacond}
2\beta\equiv \frac{\rc}{2 \muc \tc^2}\ll 1~.
\end{equation}

Let us now consider two bubbles,
one with constant Higgs phase $\theta_0$
centred at $x=y=0$, $z=\rc$
and the other with constant Higgs phase
$-\theta_0$ centred at $x=y=0$, $z=-\rc$.
It is convenient to express the Higgs modulus in the
bubbles as a product of the vacuum expectation value $\eta$ and
a function $f$ of the variables
\begin{equation}
\label{upmdef}
u_{\pm} = \frac{R_0^2 - (z \mp \rc)^2+\tau^2}{2 \rc}~,
\end{equation}
which possess O(1,3) symmetry with respect to the nucleation
events $(0,0,0,\pm
\rc)$, respectively. The modulus of the Higgs field
in each bubble is then
\begin{equation}
\label{fdef}
\eta f(u_\pm) = \phi_{\rm b}(\sqrt{x^2+y^2+(z\mp\rc)^2 -t^2}\,)=
\phi_{\rm b}(\sqrt{R_0^2 - 2 \rc u_\pm}\,)~,
\end{equation}
where $\phi_{\rm b}$ is the approximate bounce solution given by
Eq.~(\ref{bubsol1}) or, alternatively, Eq.~(\ref{rhoapprox1}).
The positions of the bubble walls
correspond to $u_\pm=0$.
The profile function $f(u_\pm)$ has non-negligible
variation only near the bubble walls or, more precisely,
for $|u_\pm|\lsim \muc^{-1}$, where
$\muc^{-1}=R_0/(\mu_0\rc)$
is the width of the bubble wall at the collision time $\tc$ given by
equation (\ref{thick1}). Using the thin-wall condition
(\ref{thincond}) it is easy to show that $\muc^{-1}\ll R_0^2/(2 \rc)$.
The argument of the bounce profile, Eq.~(\ref{fdef}), can therefore be
safely linearised, giving
\begin{equation}
\label{flin}
\eta f(u_\pm) \approx \phi_{\rm b}(R_0 - \rc u_\pm/R_0)~.
\end{equation}

The linear superposition (\ref{linsup}) can now be written
\begin{equation}
\label{linsup2}
\Phi(x)=X(x) e^{i\tht(x)} = \eta\left[f(\upl) e^{i\theta_0} +
f(\umi) e^{-i\theta_0}\right]~.
\end{equation}
We obtain the following expressions, which are assumed to be valid for
$\tau\leq \tc$:
\begin{eqnarray}
\label{Xexpr}
X^2 &=& \eta^2 \left\{ [f(\upl)+f(\umi)]^2\cos^2\theta_0 +
[f(\upl)-f(\umi)]^2\sin^2\theta_0 \right\}~,\\*
\label{theta}
\tht&=& \arctan\left[\tan\theta_0
\frac{f(\upl)-f(\umi)}{f(\upl)+f(\umi)}\right]~,\\*
\label{dtautheta}
\partial_\tau\tht&=& \sin 2\theta_0\,
\frac{\tau}{\rc}\frac{\eta^2}{X^2}
\left[ f'(\upl) f(\umi) - f'(\umi)f(\upl)\right]~,\\*
\label{dztheta}
\partial_z\tht&=& \sin 2\theta_0\,
\frac{\eta^2}{X^2}
\left[\left(1-\frac{z}{\rc}\right)f'(\upl) f(\umi)+ \left(1 +
\frac{z}{\rc}\right)f'(\umi)f(\upl)\right]~,
\end{eqnarray}
{\samepage where $\tht$ is the phase of the Higgs field in the
temporal gauge.}

At the collision the
argument $u_\pm$ simplifies to
\begin{equation}
\label{zattc}
u_{\pm}= \pm z \left(1\mp \frac{z}{2\rc}\right)~,
\quad \tau=\tc~.
\end{equation}
We see that $u_\pm$ is zero not only in the plane of collision $z=0$
but also on the two remote bubble walls located at $z=\pm 2\rc$.
However, no interesting dynamics takes place near the remote bubble
walls, and we can ignore their presence.
This is because physical fields excited by the collision can extend at
most
a distance
 $\sim\muc^{-1}$
outside the
null surface $I^0({\rm C})$ drawn in Fig.~\ref{drawing}. The
protrusion
beyond the ideal null surface
is due to the finite thickness of the bubble wall.
It can be shown that the distance between $I^0({\rm C})$
and the remote wall remains always larger than $\rc + \tc$. Therefore,
because of the no-overlap condition (\ref{nooverlap}) and the
thin-wall condition (\ref{thincond}),
physically interesting fields will never reach the remote bubble
walls.

Concentrating instead on the collision region near $z=0$, we notice
that, for $\tau=\tc$, the profile function $f(u_\pm)$ has non-negligible
variation only for $|z|\lsim \muc^{-1}$ and quickly approaches
a constant value for larger $|z|$. From Eq.~(\ref{nooverlap})
we find that $\muc^{-1}\ll 2 \rc$ and we can thus
ignore the term of order $z/\rc$ in Eq.~(\ref{zattc}),
obtaining
\begin{equation}
\label{zattcapprox}
u_{\pm}\approx \pm z~,
\quad \tau=\tc~.
\end{equation}
Similarly, for $\tau\leq\tc$ the products $f'(u_\pm)f(u_\mp)$ in
Eq.~(\ref{dztheta}) are
non-zero only for $|z|\lsim \muc^{-1}$, and we can neglect the
term $z/\rc$ also in this equation, which reduces to
\begin{equation}
\label{dzthetaapprox}
\partial_z\tht= \sin 2\theta_0\,
\frac{\eta^2}{X^2}
\left[f'(\upl) f(\umi)+ f'(\umi)f(\upl)\right]~,\quad\quad \tau\leq\tc~.
\end{equation}

\subsection{Solution for the Abelian Field Strength}

In this section, we show that the initial evolution of the scalar Higgs
field in the linear superposition approximation generates non-zero
vector fields already before the collision time $\tc$ from the
exponential tails preceeding the expanding bubble walls.
The vector field is evolved to $\tau=\tc$, and its
values there are used as initial conditions for the
ensuing linear evolution.

For non-constant Higgs modulus $X$,
the $z$ component of the Maxwell field
equation,
Eq.~(\ref{fstordereq}),
generalises to
\begin{equation}
\label{a1eqaxial}
\partial_z (3 a +\tau\partial_\tau a)=2 e X^2 \partial_z\tha~,
\end{equation}
where $\tha$ is the phase of the Higgs field in the axial gauge
defined by $A^3=0$. This phase can be expressed in terms
of $\tht$, the phase in the temporal gauge defined by $A^0=0$,
as follows
\begin{equation}
\label{gaugetrans}
\tha=\tht - \frac{e}{2} \int_{-\infty}^{\tau^2} a\, d(\tau^2)~,
\end{equation}
where we used the initial condition that $A^\mu\to 0$ and
therefore $\tha-\tht\to 0$ as $\tau^2=t^2-\rho^2\to-\infty$. One
obtains the
differential-integral equation
\begin{equation}
\label{a1eqtemp}
\partial_z(3a+\tau\partial_\tau a)=2 e X^2 \left(\partial_z\tht
- \frac{e}{2}\int_{-\infty}^{\tau^2}\partial_z a\, d(\tau^2) \right)~.
\end{equation}
We shall show at the end of this section
that the integral term on the right-hand side of this
equation is
subdominant and can be neglected provided that
\begin{equation}
\label{noacond}
\beta^2 m^2\tc^2\cos^2\theta_0\frac{\pi^2}{12}\ll 1~.
\end{equation}
Using Eq.~(\ref{dzthetaapprox})
the equation for $a$ then becomes
\begin{equation}
\label{a1eqexpl}
\partial_\tau (\tau^3\partial_z a)=\frac{m^2}{e}\sin 2\theta_0
\,\,
\tau\, \partial_\tau\! \left[f(\upl) f(\umi)\right]~.
\end{equation}

We now specialise to the approximation of the bounce solution given by
Eq.~(\ref{rhoapprox1}). Using Eq.~(\ref{flin}) one obtains
\begin{equation}
\label{ffunc}
f(u)=\frac{1}{2}[1 + \tanh(2\muc u)]~.
\end{equation}
Equation (\ref{a1eqexpl}) can then be integrated provided that we
make another well founded approximation. Realising that the composite
function $(f\circ u)(\tau)$ is monotonically increasing for all $z$,
we have for $\tau\leq \tc$ the leading behaviour $f(u)\sim g(z)
\exp[4\muc\tc(\tau-\tc)/\rc]$ for some function $g(z)$. The
right-hand side of Eq.~(\ref{a1eqexpl}) consequently
has the leading behaviour
$\sim h(z)\tau^2 \exp[8\muc\tc(\tau-\tc)/\rc]$ for some function
$h(z)$. Its integral with respect to $\tau$ is thus dominated by
contributions from an interval $\tc-\delta\lsim\tau\leq\tc$, where
$\delta=\rc/(8\muc\tc)=\beta\tc/2\ll \tc$.
Let us now consider the identity
$\tau^2 = \tc^2 + 2 \tc (\tau-\tc)[1 + (\tau-\tc)/(2\tc)]$. Near
$\tau=\tc$ we can replace $\tau^2$ by its linear expansion, provided
that $|\tau-\tc|/(2\tc)\ll 1$. However, the latter condition
is always satisfied
in the interval $(\tc-\delta,\tc]$ by virtue of Eq.~(\ref{betacond}).
We can therefore consistently replace the variables $u_\pm$
by new variables $v_\pm$, linear in $\tau$, defined by
\begin{equation}
\label{vdef}
v_\pm = \pm z + \frac{\tc}{\rc}(\tau-\tc)~.
\end{equation}
A term $-z^2/(2\rc)$ was omitted from $v_\pm$ in analogy with the
discussion
leading to Eq.~(\ref{zattcapprox}). To justify this, let us again
ignore the
remote bubble walls located at $z=\pm(\rc+\sqrt{\rc^2+\tau^2-\tc^2})$
and take into account that $f(v_\pm)$ has non-negligible variation
only for $|v_\pm|\lsim\muc^{-1}$. It then follows from the triangle
inequality and Eq.~(\ref{nooverlap}) that $|z|\leq
|v_\pm|+\tc|\tau-\tc|/\rc \lsim 9\muc^{-1}/8 \ll 2 \rc$ in the
interval $\tc-\delta<\tau\leq \tc$.

The linearisation (\ref{vdef})
has the added benefit that $f(v_\pm)$ has an exponential fall-off
as \mbox{$\tau\to -\infty$} (in contrast, $f(u_\pm)$ approaches unity as
$\tau\to -\infty$ and falls off exponentially only for $\tau^2\to
-\infty$).
Because the contribution
from $\tau\ll\tc-\delta$ is negligible, for the purpose of integration
we can extend the domain of definition
of the fields to $\tau\in(-\infty,\tc]$, identifying only
positive $\tau$ with
the physical quantity
$\sqrt{t^2-\rho^2}$.

By means of integration by parts, and using the initial condition
$a(\tau\to-\infty,z)=0$, the solution of Eq.~(\ref{a1eqexpl}) is written
\begin{eqnarray}
\label{dzasol}
\partial_z a &=& \frac{m^2}{e}\frac{\rc}{\tau^3} \sin 2\theta_0 \left\{
\tau f(\vpl) f(\vmi) - \int_{-\infty}^{\tau} f(\vpl)f(\vmi)
d\tau\right\}\nonumber\\*
&=&
\frac{m^2}{e}\frac{\rc}{4\tau^3}\sin 2\theta_0 \Big\{\tau
(1 + \tanh 2\muc\vpl)(1 + \tanh 2\muc\vmi)\nonumber\\*
&&{}-
2\beta\tc \left[ \ln(1+e^{4\muc v_{\scr +}}) +  \ln(1+e^{4\muc
v_{\scr -}})\right]\nonumber\\*
&&{}+2\beta\tc \left[ \ln(1+e^{4\muc v_{\scr +}}) -
\ln(1+e^{4\muc v_{\scr -}})\right]\coth 4\muc z\Big\}~,
\end{eqnarray}
where $\beta$ is defined in Eq.~(\ref{betacond}). The expression
for $\partial_z a$ can be integrated, using the
condition $a(\tau,0)=0$ prescribed by symmetry. One
obtains
\begin{eqnarray}
\label{li2a}
a(\tau,z)&=&\beta \frac{\tc^2}{\tau^3}\frac{m^2}{e}\sin 2\theta_0
\left\{ \frac{\tau b^2}{b^2-1}\ln\frac{1 + b c}{b+c}
\right.\nonumber\\*
&&+\frac{1}{2}\beta\tc\left[
\ln(c)\ln(1-b^2) +
\li{2}\left(-\frac{b(1+c)}{c(1-b)}\right) +
\li{2}\left(\frac{b(c-1)}{c(1+b)}\right) -
2\li{2}\left(-\frac{b}{c}\right)\right.\nonumber\\*
&&\left.\left.\quad\quad -\li{2}\left(-\frac{b(1+c)}{1-b}\right) -
\li{2}\left(\frac{b(1-c)}{1+b}\right) +2\li{2}(-bc)
\right]
\right\}~,
\end{eqnarray}
where
\begin{equation}
\label{bcdef}
b=\exp\left[4\muc \frac{\tc}{\rc}(\tau-\tc)\right]~,\quad\quad
c=\exp(4 \muc z)~,
\end{equation}
and where the poly-logarithm function $\li{n}(z)$ is defined by
\cite{deWit,lewin}
\begin{equation}
\label{lidef}
\li{n}(z)=\frac{(-1)^n}{(n-1)!}\int_0^1 dx \frac{\ln^{n-1}x}{x-1/z}
=\sum_{k=1}^{\infty} \frac{z^k}{k^n}~.
\end{equation}
For real $z$ and $n=2$
the power series is convergent for $|z|\leq 1$.
The evaluation of $\li{2}(z)$ for $z$ near or outside
the convergence radius of the series is discussed in
Appendix B.

The expressions (\ref{dzasol}) and (\ref{li2a}) are valid
for $\tc-\delta\lsim\tau\leq\tc$ and approach zero rapidly for
$\tau<\tc-\delta$. Nevertheless, these expressions for
$\partial_z a$ and $a$ have poles at $\tau=0$.
The poles are an artifact of the linearisation (\ref{vdef})
and would cancel in
the exact solutions, leading to finite (and exponentially
small) functional values of $a$ and $\partial_z a$ as $\tau\to 0$.
Because $\delta\ll\tc$ one
can easily remedy this slight problem by explicitly
setting $a(\tau,z)\equiv 0$ for $\tau\ll\tc-\delta$.

As the next step, we shall now find expressions for $a(\tc,z)$ and
$\partial_\tau a(\tc,z)$ to use as initial conditions for the linear
equations governing the evolution for $\tau>\tc$.
Taking the limit $\tau\to\tc$ in Eq.~(\ref{li2a}), using the
asymptotic result (\ref{li2asymp}) for the dilogarithm $\li{2}(x)$
(see Appendix B)
as well as the relations (\ref{recrel}) and (\ref{minrel}),
we find
\begin{eqnarray}
\label{atc}
\lefteqn{a(\tc,z)=
\label{atc2}
\beta\,\frac{m^2}{e}
\frac{\sin 2\theta_0}{2}\left\{\rule{0pt}{6mm}
\tanh 2\muc z
+ \beta
\left[\rule{0pt}{4mm} 8\muc z\ln(2\cosh 2\muc z)\right.\right.}\nonumber\\*
&&+\left.
\li{2}(\tanh 2\muc z)
-\li{2}(-\tanh 2\muc z)
+ 2\li{2}(- e^{4\muc z})
- 2\li{2}(-e^{-4\muc z})\left.
\rule{0pt}{4mm}\right]\rule{0pt}{6mm}\right\}\\*
\label{atc3}
&\approx & \beta\left(1-\beta\frac{\pi^2}{12}\right)
\frac{m^2}{e}\frac{\sin 2\theta_0}{2}
\tanh 2\muc z~.
\end{eqnarray}
The expression (\ref{atc3}) is an excellent
approximation; it was obtained by matching the functions in the limit of
large $|z|$ using Eq.~(\ref{li2asymp}) and the special values
 $\li{2}(0)=0$, $\li{2}(-1)=-\pi^2/12$ and
$\li{2}(1)=\pi^2/6$.

Similarly, by differentiating Eq.~(\ref{li2a}) with respect to $\tau$
and taking the limit
$\tau\to\tc$, we obtain
\begin{eqnarray}
\label{dtauatc1}
\partial_\tau a(\tc,z)&=& -\frac{3}{\tc} a(\tc,z) + \frac{m^2}{e\tc}
\frac{\sin 2\theta_0}{2}\tanh 2\muc z\nonumber\\*
&\approx &\left(1-3\beta+\frac{\pi^2}{4}\beta^2 \right)
\frac{m^2}{e\tc}
\frac{\sin 2\theta_0}{2}\tanh 2\muc z~.
\end{eqnarray}
Note that, in the limit of an infinitely
steep bubble wall
($\muc\to\infty$, $\beta\to 0$) and for
small Higgs phase difference $2 \theta_0$,
Eqs.~(\ref{atc}) and (\ref{dtauatc1})
reduce to the initial conditions
$a(\tc,z)=0$, $\partial_\tau a(\tc,z)=m^2\theta_0\,{\rm sgn}(z)/(e\tc)$
of the Kibble-Vilenkin step function approximation (see Sec.~\ref{step}
and Ref.~\cite{KibVil}).

The solution of Eq.~(\ref{alineq}) for $\tau\geq \tc$ subject to the
initial conditions (\ref{atc3}) and (\ref{dtauatc1}) is expressed
as
\begin{eqnarray}
\label{linsupa}
a(\tau,z)&=& \frac{m^2}{e} \frac{\sin 2\theta_0}{8\muc}
\frac{\tc}{\tau^3}
\int_{-\infty}^\infty dk
\frac{\sin k z}{\sinh \frac{\pi k}{4\muc}}\left\{\left[
-\frac{\tau-\tc}{\omega^2\tc} +\beta(1 - \beta\frac{\pi^2}{12})\tc\tau\right]
\cos\omega(\tau-\tc)\right.\nonumber\\*
&&\hspace*{4cm}\left. +
\left[\frac{\tau}{\omega}+\frac{1}{\omega^3\tc}
-\beta(1-\beta\frac{\pi^2}{12})\frac{\tc}{\omega}\right]
\sin\omega(\tau-\tc)\right\}~,
\end{eqnarray}
where $\omega^2=m^2 + k^2$.
We have retained as many orders of the small parameter $\beta$ as
is necessary to
ensure that this solution joins smoothly with the
solution for $\tau\leq\tc$ given in Eq.~(\ref{li2a}).

The Abelian field strength is
defined by $B^\varphi = \rho \partial_z a$, where $\rho=\sqrt{x^2+y^2}$
and where $\partial_z a(\tau,z)$ is given by Eq.~(\ref{dzasol}) for
$\tau\leq\tc$, or obtained for
$\tau\geq\tc$  by differentiating Eq.~(\ref{linsupa}).

\subsection{Solution for the Gauge-Invariant Phase}

Let us now turn to the gauge-invariant phase $\tilde{\theta}$.
Because of Eq.~(\ref{giphase2}) it is identical to the phase $\tha$
in the axial gauge defined by $A^3=0$. The phase $\tha$
is given by Eq.~(\ref{gaugetrans})
which has two terms, one being the phase $\tht$ in the temporal gauge
and the other an integral of the vector-potential field $a(\tau,z)$.
For $\tau\leq\tc$, these can be evaluated using the linear
superposition approximation.

The phase $\tht$ is
given explicitly
by Eq.~(\ref{theta}). For $\tau<\tc$ we shall therefore
use the more complicated, but precise,
expressions
for the bubble profile function with $f(u)$ given by Eq.~(\ref{fdef})
and $\phi_{\rm b}$ given by Eq.~(\ref{bubsol1}). This bounce solution
$\phi_{\rm b}$ has the correct exponential tail far from the bubble
wall, which is particularly important for the global behaviour of
gauge-invariant phase $\tilde{\theta}$.

The integral in Eq.~(\ref{gaugetrans}) can be evaluated using the
expression for $a(\tau,z)$ given in Eq.~(\ref{li2a}), provided that
one expands inverse powers of $\tau$ in a Taylor series about $\tau=\tc$.
It suffices to do the calculation
to lowest order in the parameter $\beta$.
One finds
\begin{eqnarray}
\label{atauint}
\lefteqn{e\int_{-\infty}^{\tau} d\tau\,\tau a = \beta^2 m^2\tc^2
\frac{\sin 2\theta_0}{2}\ \times} \nonumber\\*
&&{}\times\left\{
\frac{1}{2}\ln\frac{b+c}{1+bc}\ln\!\left(\frac{(1+bc)\left(1+\frac{b}{c}
\right)\left(\sqrt{c}+\frac{1}{\sqrt{c}}\right)^2}{(1-b^2)^2}\right) +
\ln c\ \ln\! \left(\frac{(1+bc)\left(1+\frac{b}{c}\right)}%
{\sqrt{c}+\frac{1}{\sqrt{c}}}\right)
\right.\nonumber\\*
&&\quad{}+\li{2}(1-c)-\li{2}\left(\frac{c-1}{c}\right)-
\li{2}\left(\frac{1}{1+c}\right) +
\li{2}\left(\frac{c}{1+c}\right)\nonumber\\*
&&\left.\quad{}+\li{2}\left(\frac{c-1}{b+c}\right) -
\li{2}\left(\frac{b+c}{1+c}
\right)
-\li{2}\left(\frac{1-c}{1+bc}\right) +
\li{2}\left(\frac{1+bc}{1+c}\right)
\rule{0pt}{8mm}\right\}~,
\end{eqnarray}
which is valid for $\tau\leq\tc$.
In particular, at $\tau=\tc$ the expression reduces to
\begin{eqnarray}
\label{atauinttc}
e\int_{-\infty}^{\tc} d\tau\,\tau a\!\!\!&=&\!\!\! \beta^2 m^2\tc^2
\frac{\sin 2\theta_0}{2}\left\{\rule{0pt}{5mm} 4 \muc z
\ln(2\cosh 2\muc z)+
\li{2}(\tanh 2\muc z)-\li{2}(-\tanh 2\muc z)\right.\nonumber\\*
&+&\!\!\!\!\left.\li{2}(1-e^{4\muc z})-\li{2}(1-e^{-4\muc z}))-
\li{2}\left(\frac{1}{1+e^{4\muc z}}\right) +
\li{2}\left(\frac{1}{1+e^{-4\muc z}}\right)\rule{0pt}{5mm}\right\}
\nonumber\\*
&\approx& \beta^2 m^2\tc^2
\frac{\sin 2\theta_0}{2}\frac{\pi^2}{12}\tanh 2\muc z~.
\end{eqnarray}

We are now ready to derive approximative
expressions for $\tha(\tc,z)$ and
$\partial_\tau\tha(\tc,z)$ to be used as initial conditions for the
ensuing linear evolution. First, from Eqs.~(\ref{theta})
and (\ref{zattcapprox}) we obtain
\begin{eqnarray}
\label{thetaapprox}
\tht(\tc,z)&=&\arctan\left[\tan\theta_0\frac{f(z)-f(-z)}{f(z)+f(-z)}\right]
\nonumber\\*
&\approx&\arctan(\tan\theta_0 \tanh 2\muc z)\approx \theta_0\tanh\left[
\frac{\tan\theta_0}{\theta_0} 2\muc z\right]~,
\end{eqnarray}
where we used Eq.~(\ref{ffunc}) in the intermediate step. Substituting
this result together with Eq.~(\ref{atauinttc}) into Eq.~(\ref{gaugetrans})
we find the initial condition
\begin{equation}
\label{thatc}
\tha(\tc,z) \approx \theta_0\tanh\left[
\frac{\tan\theta_0}{\theta_0} 2\muc z\right] -
 \beta^2 m^2\tc^2
\frac{\sin 2\theta_0}{2}\frac{\pi^2}{12}\tanh 2\muc z~.
\end{equation}
Turning next to the $\tau$ derivative given by Eq.~(\ref{dtautheta}),
using
Eqs.~(\ref{zattcapprox}) and (\ref{ffunc})
we have that
\begin{equation}
\label{dtauthetatc}
\partial_\tau\tht(\tc,z) = - \frac{2\muc\tc}{\rc}\tan\theta_0\,
g(\muc z)~,
\end{equation}
where the function $g(x)$ is defined by
\begin{equation}
\label{gdef}
g(x)=\frac{\tanh 2x}{\cosh^2 2x +\tan^2\theta_0\sinh^2 2x}~.
\end{equation}
We shall need an approximation of the function $g$ which possesses a
simple Fourier transform. To this effect,
let us define the value $x_0$ by
\begin{equation}
\label{z0def}
x_0=\frac{1}{4}\cosh^{-1}\left[\frac{1}{2}
(1+\sqrt{5+4\cos 2\theta_0})\right]
\end{equation}
satisfying
\begin{equation}
\label{gz0def}
g(x_0)=4\left(\frac{\sqrt{5+4\cos 2\theta_0}-1}%
{\sqrt{5+4\cos 2\theta_0}+3}\right)^\frac{1}{2}\frac{\cos^2\theta_0}%
{\sqrt{5+4\cos 2\theta_0} + 1 + 2\cos 2\theta_0}
\end{equation}
An excellent approximation to $g(x)$ is then given by
\begin{equation}
\label{gapprox}
g(x)\approx 16\,x_0^3\, g(x_0) \frac{x}{(3x_0^2+x^2)^2}~.
\end{equation}
Combining Eqs.~(\ref{dtauthetatc}), (\ref{gapprox}), (\ref{atc3})
and (\ref{gaugetrans}) one finds the other initial condition
\begin{eqnarray}
\label{dtauthatc}
\lefteqn{\partial_\tau\tha(\tc,z)\equiv\partial_\tau\tht(\tc,z)
-e\tc a(\tc,z)
}\nonumber\\*
&\approx&\displaystyle
-\frac{
32
\muc\tc}{\rc}\tan\theta_0\,\frac{x_0^3\,g(x_0)\,\muc z}%
{(3x_0^2+\muc^2 z^2)^2}
-\beta \left(1-\beta\frac{\pi^2}{12}\right) m^2\tc \frac{\sin
2\theta_0}{2}
\tanh 2\muc z~.
\end{eqnarray}
The solution of Eq.~(\ref{thetalineq})
for $\tau\geq \tc$ subject to the
initial conditions (\ref{thatc}) and (\ref{dtauthatc}) is given
by
\begin{eqnarray}
\label{thasol}
\lefteqn{\tha(\tau,z)=\frac{\tc}{\tau}\int_{-\infty}^{\infty}dk\,
\frac{\sin kz}{4\muc}
{}~\times}\nonumber\\*\left\{\rule{0pt}{7mm}\right.&&\hspace*{-9mm}
\left[\frac{\theta_0^2}{\tan\theta_0\sinh\left(
\frac{\theta_0}{\tan\theta_0}\frac{\pi k}{4\muc}\right)}-
\frac{\sin 2\theta_0}{2}\frac{\beta^2 \pi^2 m^2 \tc^2}
{12 \sinh\left(\frac{\pi k}{4\muc}\right)}\right]
\cos\,\omega(\tau-\tc)\nonumber\\*
&+&\left[\frac{\theta_0^2}{\tan\theta_0\sinh\left(
\frac{\theta_0}{\tan\theta_0}\frac{\pi k}{4\muc}\right)}-
\frac{\sin 2\theta_0}{2}\frac{\beta m^2 \tc^2}
{\sinh\left(\frac{\pi k}{4\muc}\right)}\right.\nonumber\\*
&&\left.\left.\hspace*{1cm}-\tan\theta_0\frac{32\tc^2}%
{\sqrt{3}\rc}g(x_0) x_0^2 k \exp(-\sqrt{3}x_0|k|/\muc)\right]
\frac{\sin\,\omega(\tau-\tc)}{\omega\tc}\rule{0pt}{7mm}\right\}~,\quad
\end{eqnarray}
where $\omega^2 = m^2 + k^2$.

Finally, let us justify the approximation made in
Eq.~(\ref{a1eqtemp}),
where the integral term was neglected. To
estimate the relative significance of the two terms in the right-hand
side of that equation, recall that
the integral $\int d\tau\tau \partial_z a$ is an increasing function
of $\tau$ which is significantly nonzero
only in a narrow interval $\tc-\delta\lsim\tau\leq \tc$, where
$\delta=\rc/(8\muc\tc)=\beta\tc/2 \ll \tc$. The largest values of the
integral occur at $\tau=\tc$, where it is easily compared to the other
term. From Eq.~(\ref{atauinttc}) one obtains
\begin{equation}
\label{aintestim}
e\int_{-\infty}^{\tc} d\tau \tau \partial_z a \approx \beta^2 m^2
\tc^2
\muc \sin 2\theta_0 \frac{\pi^2}{12}\,\sech^2 2\muc z~,
\end{equation}
and from Eq.~(\ref{thetaapprox}), similarly,
\begin{equation}
\label{dzthetaestim}
\partial_z\tht(\tc,z) \approx 2\muc \tan\theta_0\, \sech^2\!\left(
\frac{\tan\theta_0}{\theta_0} 2 \muc z\right)~.
\end{equation}
Therefore, the condition for the
integral in Eq.~(\ref{a1eqtemp}) to be negligible is given by
Eq.~(\ref{noacond}).

\section{Lattice Computation}
\label{latsec}

In order to compare our approximate
analytical solutions to the
true evolution of fields in the original field theory,
we have evolved the system of two bubbles numerically on the lattice,
using the temporal gauge $A^0=0$ and the techniques described in Appendix A.
As the first step of such a
lattice simulation
it is necessary
to specify the initial conditions.
In particular, one must determine the initial shape of the bubble walls.

For definiteness we consider the Higgs potential (\ref{pot})
with the fields and coordinates rescaled
so that $\lam=\eta=1$.
We choose $\epsilon=0.2$, corresponding to the thin-wall
condition (\ref{thincond}) being marginally satisfied. This means that
the bubble wall is
thin, but not extremely thin,
compared to the nucleation radius $R_0$ (see Fig.~\ref{bouncefig}).
The gauge coupling of the
rescaled fields is taken to be $e=0.5$.

The Euclidean field equation relevant to the nucleation of a single
spherical bubble at rest,
Eq.~(\ref{eucleq}),
has been solved using a shooting method.
At $t=0$ in Minkowski space the spatial dependence of the
single-bubble configuration coincides with
that of the Euclidean instanton solution.
The initial two-bubble configuration is approximated by the linear
superposition of the scalar fields of two well separated single
bubbles. The vector potential $A^i$ and the time derivatives of all
fields are set to zero initially.

We have used a spatial Cartesian grid consisting of
$125^3$ unit cells with a lattice spacing of $a=0.2$ (this $a$ need not
be confused with the field $a(\tau,z)$ defined in
Eq.~(\ref{thetaa})).
In order to maximise the dynamical range, we placed the bubbles in
two adjacent corners of our box
and employed reflective
boundary conditions. More precisely, on each face of the box we
imposed $\partial\Phi/\partial x^\perp=0$, $A^\perp=0$, and $\partial
A^\parallel/\partial x^\perp=0$,
where $\perp$ signifies the outward normal and
$A^\parallel$ denotes the two vector-potential
components parallel to the face.
Derivatives were approximated by symmetric differences, which caused the
planes of symmetry in each direction to be
located at the lattice points with indices 2 and 124. Thus, the
distance between bubble centres is $2\rc = 122 a = 24.4$.

Finally, the system was evolved using a time step of $h a$ with
$h=0.2$. The lattice Gauss constraint, Eq.~(\ref{gauss}),
was monitored and found to remain zero at the level of machine
accuracy, i.e., one part in $10^{15}$.

\section{Analysis}
\label{disc}

We shall here present results for the Abelian field strength
$B^\varphi$ and the gauge-invariant Higgs phase $\tilde{\theta}=
\int dx^\mu D_\mu\theta$ in the collision of two expanding bubbles.
Analytical results, within the Kibble-Vilenkin step-function
approximation as well as our new approach, are
compared to
the true field evolution obtained from the numerical simulation on the
lattice.

First, let us review the values of various parameters that occur in
the solutions. As was mentioned in
the last section, we take $\lam=\eta=1$
and $\epsilon=0.2$ as parameters of the Higgs potential. The gauge
coupling is $e=0.5$ and the bubble radius at the time of collision is
$\rc=12.2$~. We then find directly the vector-boson mass
$m=\sqrt{2} e\eta= 0.7071$~.
Expressions (\ref{mu0def}) and (\ref{R0value}) determine,
within the thin-wall approximation,
the initial bubble-wall slope $\mu_0=0.375$ and the nucleation
radius $R_0=7.5$~. The time of collision is obtained from
Eq.~(\ref{o31sol}) with $\ri=R_0$, which gives
$\tc=\sqrt{\rc^2-R_0^2}=
9.62$~. The bubble-wall slope at the time of collision
$\muc = \mu_0 \rc/R_0 = 0.61$ is found from Eq.~(\ref{thick1}).
As a consequence,  $\beta\equiv\rc/(4\muc\tc^2)=0.054 \ll 1$ as required.
Finally, the constant $s_0$ occurring in Eq.~(\ref{bubsol1})
is given by $s_0=6.95$. This completes a list of parameters that
are uniquely determined by the physical model and by the chosen distance
between nucleation centres, $2\rc$. The average value of this
bubble separation can also be calculated from the model parameters
\cite{Kurki-Suonio}.

The remaining parameter, $\theta_0$, is
determined by the random nature of
spontaneous symmetry
breaking. We choose $\theta_0=\pi/8$, corresponding to a particular
case in which two colliding bubbles have a Higgs phase difference of
$\pi/4$. This phase difference is sufficiently small that important
approximations remain valid, such as the assumed constancy of
the Higgs-field modulus inside the bubble overlap region
after collision.
Moreover,
it can be checked that the no-overlap condition (\ref{nooverlap})
is satisfied with the left-hand side being approximately $3.5$~; this
ensures that the overlap of two bubbles is negligible at the time
of nucleation $t=0$. Finally, the condition (\ref{noacond}),
used in the derivation of the analytical solution for
$\tilde{\theta}$, is
satisfied with the left-hand side being approximately $0.095 \ll 1$~.
With the chosen value of $\theta_0$, Eqs.~(\ref{z0def}) and (\ref{gz0def})
give $x_0=0.314$ and $g(x_0)=0.365$.

\begin{figure}[!tb]
  \begin{center}
       \leavevmode
       \epsfxsize=18cm \epsfbox[53 128 543 518]{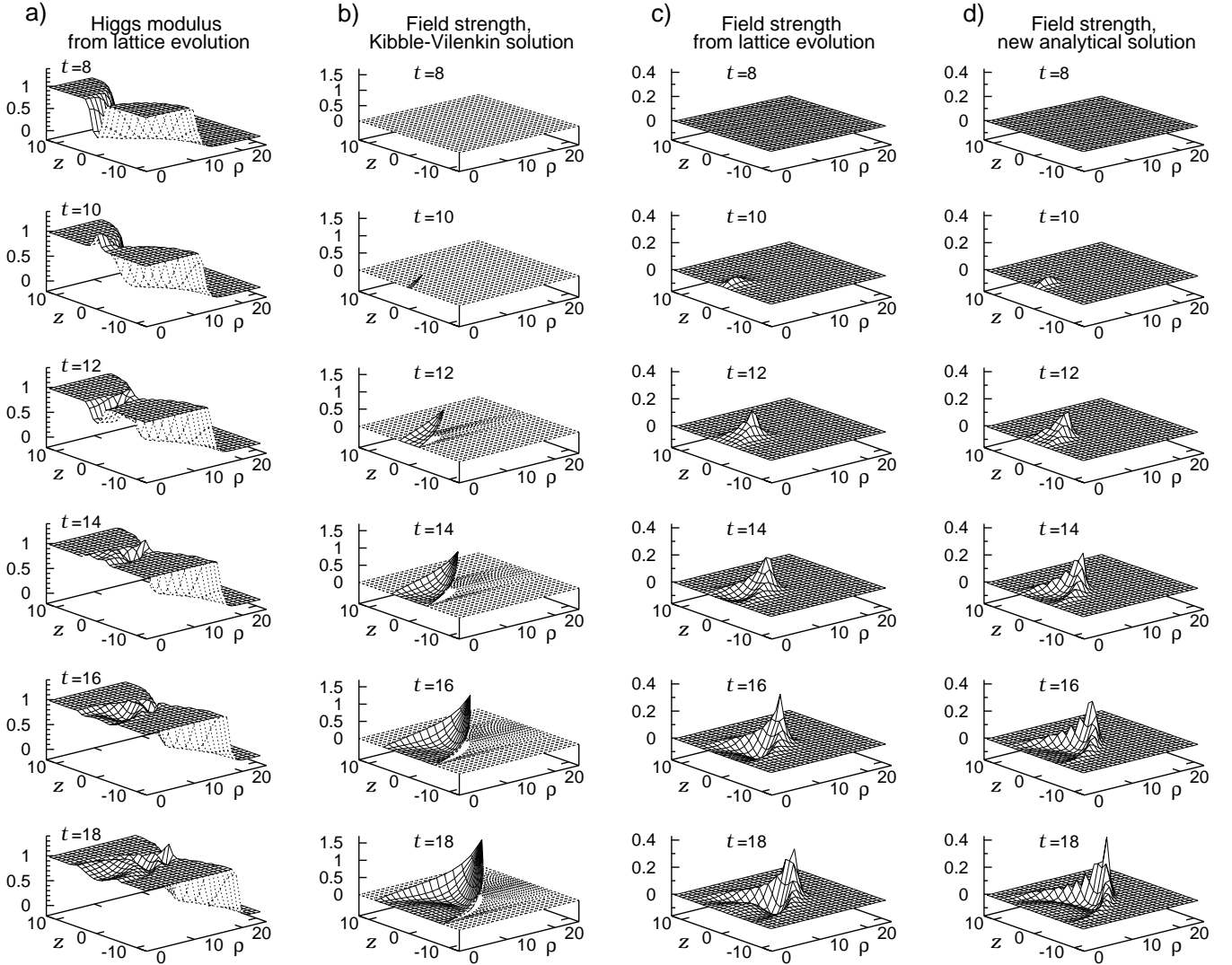}
\caption{Bubble collision with initial Higgs-field phase difference
$\Delta\theta_0=2\theta_0=\pi/4$. \mbox{a) Lattice} evolution of
the modulus $|\Phi|$ of the Higgs
field, b) Kibble-Vilenkin approximate solution for the azimuthal
field strength $B^\varphi$,
c) Lattice evolution of $B^\varphi$,
d) Our new analytical solution for $B^\varphi$.
Here $\rho=(x^2+y^2)^{1/2}$
is the
distance from the symmetry axis of the collision, $|\Phi|$ is
measured in units of $\eta$,
$B^\varphi$
in units of $m^2/e$,
and $\rho$, $z$, $t$ in units of
$\lam^{-1/2}\eta^{-1}$.
The centres of
the two bubbles are at $\rho=0$, $z=\pm 12.2$. The bubbles first
collide at time
$\tc=9.62$, which falls between the first and
second frames in each column.}
\label{bsurf}
  \end{center}
\end{figure}

The results for the Abelian field strength
are most easily explained with the help of
Fig.~\ref{bsurf}.
The first and third columns of this figure depict results from the
numerical evolution of the
fields on a lattice, which was described in
Sec.~\ref{latsec}. Because of the cylindrical symmetry, the fields
depend only on the coordinate $z$ along the symmetry axis and the
distance $\rho$ from this axis. Two bubbles were initially
nucleated
in the far left and near corners,
defined by $\rho=0$, $z=\pm 12.2$,
with nucleation radius $R_0=7.5$. In column (a) we can follow the
evolution of the modulus of the scalar field as the bubbles expand.
The bubbles first collide at time $\tc=9.62$, which falls between
the first and second frames. As they continue to expand, the
modulus within the region of overlap of the two bubbles
begins to fluctuate.
The fluctuations are moderate, and therefore
the approximation of the Higgs modulus by a constant
value $|\Phi|=\eta$ within the union of interior of the two bubbles
is well justified. A larger
distance $2\rc$ between nucleation centres would lead to higher
bubble-wall speed at collision, $v_{\rm c}=\sqrt{1-R_0^2/\rc^2}$,
and therefore to larger fluctuations in the Higgs modulus.
Violent fluctuations in the modulus are induced for large phase
differences $\Delta\theta_0\gsim \pi/2$ regardless of the collision
speed. Therefore, we don't expect our analytical treatment to remain
valid for large values of $\theta_0$.

Turning next to the field strength $B^\varphi$
depicted in column (c), we notice
that the flux of $B^\varphi$ is mainly concentrated in a peak which
moves away from the $z$ axis in the symmetry plane $z=0$.
Because of the cylindrical symmetry,
the peak corresponds to a circular flux tube enclosing the region
of bubble intersection. In fact,
by comparing
with column (a), we see that the peak is located near the value of
$\rho$ at which the bubbles have most recently come into contact.
This circle of most recent intersection is represented by
the point $C_t$ in Fig.~\ref{drawing}a and has radius
$\rho_{\rm c}=\sqrt{t^2-\tc^2}$. As a consequence, the flux tube
moves away from the symmetry axis
 with a speed greater than the speed of light. This is not
in contradiction with special relativity;
the field strength in the peak has been
generated locally and causally by the most recent bubble-wall
intersection,
while the tail toward the bubble interior
consists of fields generated by earlier intersections
occurring at smaller radius $\rho$.

In column (b) we observe that the solution obtained by Kibble and
Vilenkin using a step-function approximation (see Sec.~\ref{step}
and Ref.~\cite{KibVil}) gives a very reasonable qualitative
description of the behaviour of the field strength in the
region \mbox{$I^{+}({\rm C})_t$} $=$ \mbox{$\{(z,\rho)\,|\,
(\tc+|z|)^2 + \rho^2 < t^2\}$}, which constitutes
the causal future of bubble intersections and is shown
in Fig.~\ref{drawing}a. On the other hand, the peak value
of the field strength in this approximation is about four times higher
than the true peak value obtained from the lattice evolution.
The reason for the higher field strength is that the step-function
initial condition leads to a much higher time derivative of the field
strength at the collision than does the smooth initial condition
(\ref{dtauatc1}) obtained with finite-width bubble walls.

\begin{figure}
  \begin{center}
       \leavevmode
       \epsfxsize=4.5cm \epsfbox[61 53 172 517]{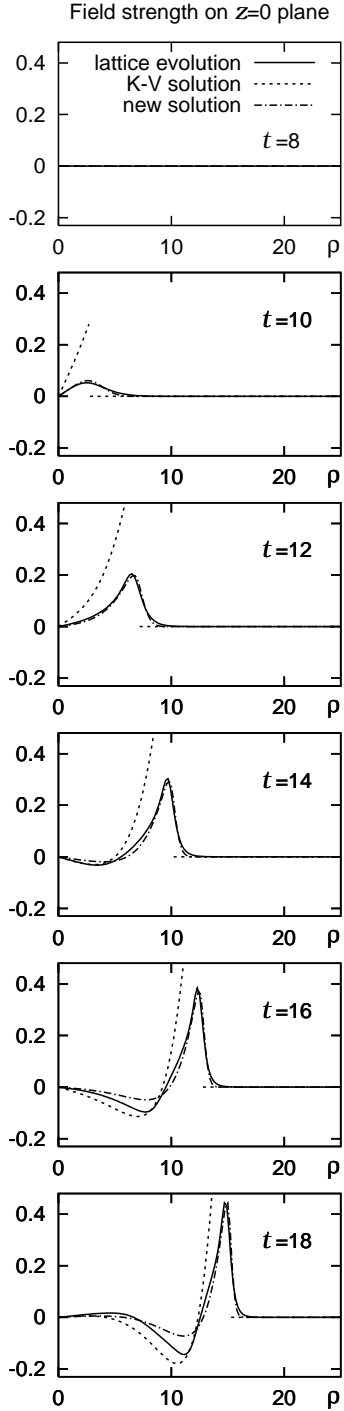}
\caption{Field strength $B^\varphi$ (in
units of $m^2/e$) on the plane of symmetry $z=0$.
The numerical solution (solid curve) is compared
with the Kibble-Vilenkin
approximate solution (dashed) and with
our new analytical solution (dot-dashed).
For $t\geq 12$,
the
Kibble-Vilenkin solution extends to higher values of $B^\varphi$
than what is visible in the graph.}
\label{bprof}
  \end{center}
\end{figure}

The new, smooth analytical solution is displayed in column (d). It is
found to be in excellent agreement with the lattice
results (c).\footnote{In
fact, the agreement is even better than it appears to be because,
in order to produce
black-and-white surface graphs such as these, one must use a sparse
mesh which cannot always be brought to
coincide precisely with peaks and other features.} The two solutions
are more easily compared in Fig.~\ref{bprof},
which shows the field strength on the symmetry
plane $z=0$. Here the agreement of the new solution with the numerical
results is rather remarkable. The tail that leads down from the
peak towards higher values of $\rho$ is accurately reproduced.
We find that the maximum field strength increases with time, and the
width of the peak $\Delta\rho$
decreases with time as predicted by the
analysis in Sec.~\ref{step}. Note that there is a trough
with negative values of the field strength trailing the primary peak
by about a peak width. This trough corresponds to an expanding
circular flux-tube loop with
smaller radius that the primary one
and with the Abelian field strength pointing in the opposite
direction. Such a secondary flux tube is created by wave-like
excitations travelling inward from previous locations of the
primary peak. In fact, the two-dimensional nature of the
bubble-collision
problem implied by the O(1,2) symmetry explains why these solutions are
similar to waves on a water surface.
The trough is less accurately reproduced by the analytical solution
as time progresses.

\begin{figure}
  \begin{center}
       \leavevmode
       \epsfxsize=9cm \epsfbox[51 127 289 522]{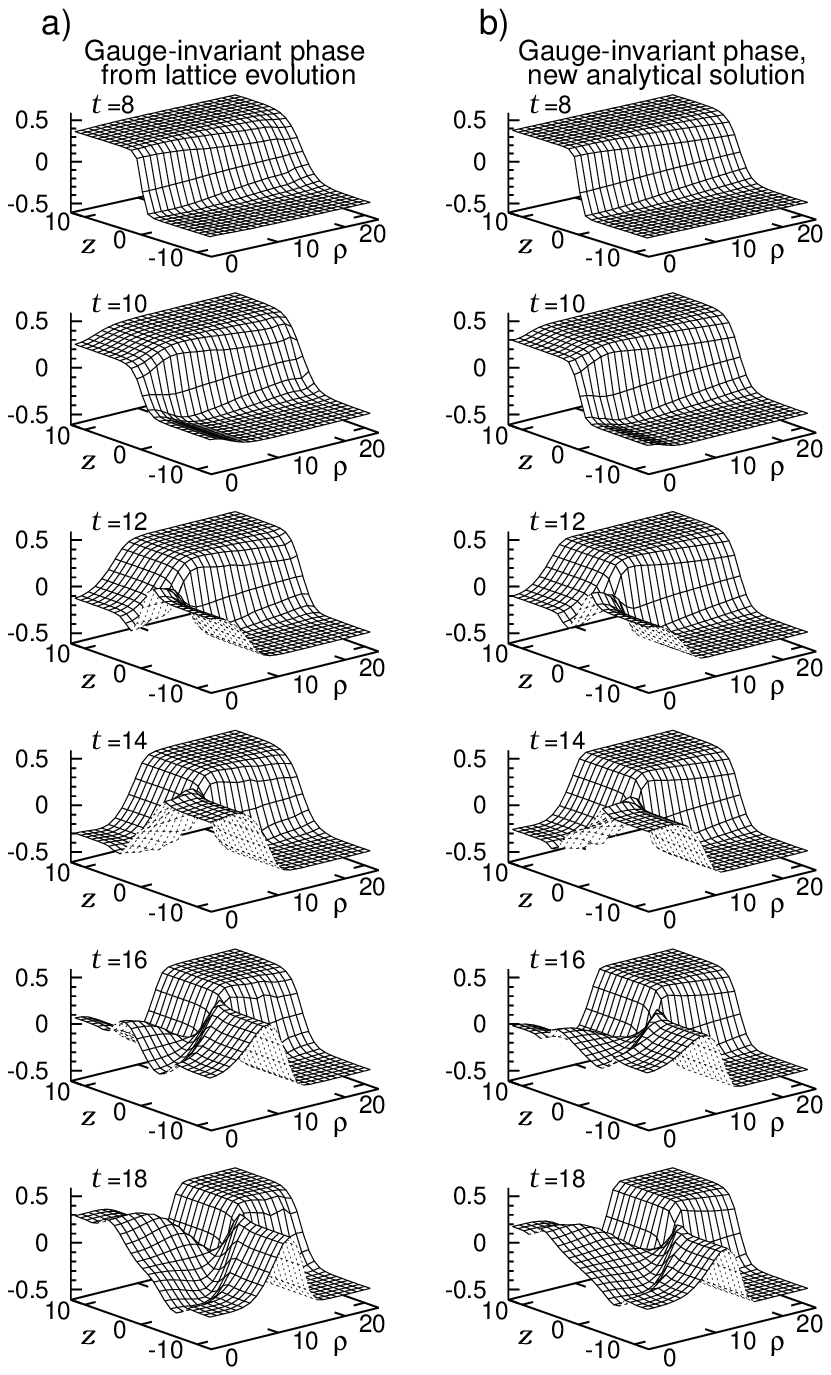}
\caption{Gauge-invariant phase  $\tilde{\theta}=\int dx^\mu
D_\mu\theta$ in a
bubble collision with initial Higgs-field phase difference
$\Delta\theta_0=2\theta_0=\pi/4$: a) Lattice evolution of
$\tilde{\theta}$, b) Our new analytical solution for $\tilde{\theta}$.
Again, $\rho=(x^2+y^2)^{1/2}$
is the distance from the symmetry axis of the collision
and $\rho$, $z$, $t$ are measured in units of
$\lam^{-1/2}\eta^{-1}$. The bubbles first
collide at time
$\tc=9.62$, which falls between the first and
second frames in each column.}
\label{githeta}
  \end{center}
\end{figure}

Let us now consider the gauge-invariant Higgs phase
$\tilde{\theta}=\int dx^\mu D_\mu\theta$, properly defined in
Eq.~(\ref{giphase2}). This phase is equal to the Higgs phase $\tha$
in the axial
gauge. Fig.~\ref{githeta} shows a comparison between
the numerical solution for the phase and
the analytical solution.
As the bubbles collide, the phase
in the interior of the bubbles
begins to oscillate (to see the
location of the bubbles at each time step, refer to Fig.~\ref{bsurf}a).
Note that
$\tilde{\theta}$
changes instantly also for
large values of $|z|$, which may at first seem unphysical.
The phase
$\tilde{\theta}$ is, however, a non-local quantity by construction,
obtained by
integrating the local gauge-invariant phase gradient $D_\mu\theta$.
One can check that $D_\mu\theta$ is non-zero only within the
causal future of bubble-wall intersections, as should be expected
from a physical quantity. Although $\tilde{\theta}$ has no direct
physical meaning,
it provides a useful, gauge-invariant characterisation
of the field configuration, by means of which different solutions can be
easily compared. It is also a very sensitive measure, because small
deviations in $D_\mu\theta$ accumulate as they are integrated.
The agreement between the numerical and analytical
solutions in Fig.~\ref{githeta} is nevertheless
quite good. Through careful study
one finds that the
analytical solution is slightly delayed in time after the collision,
and also that the amplitude of oscillation is smaller in the
analytical solution than in the numerical evolution. We believe that
these small deviations
are the result of non-linear evolution near the time of collision
that our analytical approach is unable to account for.

A particularly interesting quantity is the gauge-invariant phase
difference of the two bubbles.
For $\tau>\tc$ this is obtained by considering
the $z\to\pm\infty$ limits of Eq.~(\ref{thasol}), while
Eqs.~(\ref{theta}), (\ref{atauint}) and (\ref{gaugetrans})
in the same limits
give the result for $\tau<\tc$. With the definition
$\Delta\tilde{\theta}(\tau)\equiv \tilde{\theta}(\tau,z\to+\infty) -
\tilde{\theta}(\tau,z\to-\infty)$
we find
\begin{eqnarray}
\label{zinfhigh}
\Delta\tilde{\theta}(\tau)
&=&
\frac{2\tc}{\tau}\left[\left(\theta_0 - \frac{\sin 2\theta_0}{2}
\frac{\beta^2 \pi^2 m^2 \tc^2}{12}\right)\cos m (\tau-\tc)\right.
\nonumber\\*
 &&\quad\quad\left.{}+
\left(\theta_0 - \frac{\sin 2\theta_0}{2}
\beta  m^2 \tc^2\right)\frac{\sin
m(\tau-\tc)}{m\tc}\right]\quad\quad,~\tau>\tc
\end{eqnarray}
and
\begin{eqnarray}
\label{zinflow}
\Delta\tilde{\theta}(\tau)
&=&
2\left[\theta_0 - \frac{\sin 2\theta_0}{2}\beta^2 m^2 \tc^2 \left(
\frac{1}{2}\ln(b)\ln\frac{(1-b^2)^2}{b} +\li{2}(b) -
\li{2}(-\frac{1}{b})
-\frac{\pi^2}{6}\right)\right]\nonumber\\*&~&\hspace*{10cm},~\tau<\tc
\end{eqnarray}
where $b$ is defined in Eq.~(\ref{bcdef}).
In particular, on the axis through the bubble centres ($\rho=0$)
the phase difference
is
given by the above expressions with the substitutions $\tau\to t$ and
$b\to\tilde{b}=\exp[4\muc\tc(t-\tc)/\rc]$.

The oscillatory behaviour of the phase difference
$\Delta\tilde{\theta}(t)$
after the collision ($t>\tc$)
is fully accounted for in Eq.~(\ref{zinfhigh}).
The amplitude of oscillations is proportional to $t^{-1}$. The phase
difference between the interior of the
bubbles therefore equilibrates to zero as time
progresses.
This behaviour was first noted by Kibble and Vilenkin \cite{KibVil}.
In fact, in the limit of infinitely thin bubble walls ($\beta\to 0$),
Eq.~(\ref{zinfhigh}) agrees with Eq.~(\ref{thetanice}),
which is their result. On the other hand, the phase difference
far from the symmetry axis remains $2\theta_0$.
This is evident in Fig.~\ref{githeta}a,
but also follows analytically from Eq.~(\ref{zinflow})
in the $b\to 0$ limit, using Eq.~(\ref{recrel}) of Appendix B.

Finally,
we shall derive
the total flux of
the Abelian field strength. This is the flux
through a semi-infinite plane based on the $z$
axis, and can be computed in analogy with the discussion leading to
Eq.~(\ref{totflux}) by evaluating the line integral of the vector
potential
$A_\mu$ at a constant time $t$ along the closed path ABDEA indicated in
Fig~\ref{drawing}a. With finite-thickness bubble walls, however, it is
necessary to let the $z$ coordinates of points A and B approach $\pm
\infty$, and similarly to take the $\rho$ coordinate of points D and E
to infinity.
Proceeding as in Ref.~\cite{KibVil},
we realise that the line integral of $e A_\mu$ is equal to the line
integral of $D_\mu\theta$, since the line integral of $\partial_\mu\theta$
along a closed path vanishes. Moreover, $D_\alpha\theta$ must vanish on
the line segments BD and EA, because they are far removed from the
causal future of bubble intersections $I^{+}(C)_t$.
Note, however, that $D_z\theta$ is non-zero on the segment ED.
This is because the initial condition establishes
a phase difference
of $2\theta_0$ between D and E,
which contributes to $\partial_z\tha$.
 In the axial gauge $A_z=0$, and we thus obtain
\begin{equation}
\label{totflux2}
\Phi_B(t)=-\frac{1}{e}
\left(\int_{\rm AB}+\int_{\rm DE}\right)dx^\mu \partial_\mu\tha=
\frac{1}{e}\left(2\theta_0 - \Delta\tilde{\theta}(t)\right)~,
\end{equation}
where $\Delta\tilde{\theta}$ is given by Eq.~(\ref{zinfhigh}) or
Eq.~(\ref{zinflow}). As a function of time, the Abelian flux rises
from zero and performs damped oscillations about the late-time limit
$2\theta_0/e$.
This asymptotic value agrees with that
found for the Kibble-Vilenkin solution
in Sec.~\ref{step}.

\section{Conclusions}

In this paper, we have attempted to provide the most complete
analysis to date of the phase equilibration and field-strength
evolution that occur in a
U(1) gauge theory when two bubbles of true vacuum collide. Our
analysis has been motivated by the work of Kibble and
Vilenkin \cite{KibVil}, and we have been able to go beyond some
of their assumptions concerning the initial configuration of
the phase of the scalar field. In particular, we have found
that a linear superposition
ansatz, coupled with smooth
bubble profiles arising from bounce solutions
to the Euclidean equations of motion, leads
to analytical solutions that are
in excellent agreement with the full numerical evolution of the fields,
at least for small initial phase difference between the
two bubbles.

In Sec.~4, we presented the new, smooth analytical solutions for the
Abelian field strength and the gauge-invariant phase. They have a number of
advantages over solutions obtained in Ref.\cite{KibVil} using
the step-function approximation. For
example,
from the latter one obtains
a peak value of the field strength that differs
by a factor of four from the correct value obtained
in numerical simulations (a factor of sixteen
difference in energy density). Moreover,
the solutions (\ref{thetanice})--(\ref{covalpha}) derived from the
step-function approximation
have discontinuities on the future null surface of bubble
intersections and so are unable to account for the
behaviour of the fields near that surface. The solutions are only
valid over limited ranges of the coordinate
$\tau=\sqrt{t^2-x^2-y^2}$, which means that they
provide no information about the field dynamics
shortly before the bubble collisions or, at later times,
outside the bubble walls. The smooth ansatz that
we adopt, on the other hand,
allows us to account for the fact that
realistic bubble walls of finite width
have exponential tails that begin to overlap
long before the instant of collision. The overlapping tails initiate
the evolution of the phase of the scalar field and of the vector
potential, so that it
 is already in an advanced stage
when the collision reaches full impact.
The advantage of the smooth approximation becomes manifest in
the comparative plots of Figs.~\ref{bsurf}--\ref{githeta}.

An interesting feature emerges in Fig.~5. There appears to be
evidence for the existence of alternating positive peaks and negative
troughs in the field strength,
corresponding to expanding circular flux-tube loops with alternating
direction of the field strength.
This raises the possibility that we are seeing here
an analogy to the
suite of alternating
vortices and antivortices
found by
Digal and Srivastava \cite{digal} in the
collision region of two colliding bubbles in a global U(1) model.
Our peaks are not large enough
to form defects, because we have set the initial
conditions so that the geodesic rule is obeyed,
but it seems that our positive and negative
peaks may be an ``embryo'' for such vortex-antivortex configurations.
This is supported by the observation
that the modulus of the Higgs field
appears to have dips at the locations
of the big peak and the trailing anti-peak. In a hard collision,
the field strength would be larger and the Higgs
modulus would go down
to zero in these two places,
corresponding to the formation of topological defects.

We should discuss some of the assumptions we have had to make in
this article. We have restricted ourselves to O(1,2)-symmetric
solutions, but, as we
have argued in Sec.~\ref{sphermem}, this symmetry
provides a good description in general for all pairs of nucleated
bubbles except possibly in the very first stages of their expansion.
Throughout our analysis we have kept the
radial component of the
complex scalar field fixed in the minimum of the potential
within the union of the interior of the two bubbles.
The validity of
this approximation has been checked against the numerical simulations
(see Fig.~\ref{bsurf}) and
has been found satisfactory for the range
of initial conditions we have adopted. The constraint
placed upon us is that we are not really in a position to discuss the
collision process involving large initial phase differences, for
that is a situation where we would expect
large gradient terms to arise and
the scalar field to fluctuate wildly away from the vacuum.  At the moment,
we do not see any obvious way to approach this problem.

The work we have presented here has a number of natural extensions.
For example, one can use our approach to obtain smooth and realistic
solutions to effective field
equations that incorporate the effects of plasma conductivity, and
thereby improve on previous results of this kind
obtained with the step-function approximation \cite{kibble,enqvist}.
Alternatively, or perhaps in addition,
one can try to include the effects of frictional damping of the bubble
walls \cite{Melfo,turok,lilley}.
This unfortunately makes the problem non-relativistic and
reduces the symmetry from O(1,2) to O(2). Therefore, it is likely that
a new type of solution, or even approximation, will be needed.

By going beyond the electroweak Standard Model, we could begin
discussing the generation of primordial magnetic fields in a first-order
transition. We have good estimates for the magnitude and width of the
magnetic fields formed from our smooth ansatz. This could be useful
in determining the possible success of generating
galactic-strength fields from such seeds in the early Universe.

\subsection*{Acknowledgments}
\setcounter{equation}{0}
\renewcommand{\theequation}{A.\arabic{equation}}
O.T. thanks M.J. Lilley
and P.M.S. thanks G. Moore for useful discussions. O.T. is grateful
to the Yale Center for Theoretical Physics for kind hospitality.
This work was supported by the
European Commission's TMR programme under Contract
No.~ERBFMBI-CT97-2697 and the
U.K.\ EPSRC
under Grant GR/K50641. E.J.C and P.M.S. were supported by the U.K.\ PPARC.
Some of this work was performed on Origin 200 machines, supported by
Silicon Graphics/Cray Research, HEFCE and PPARC.

\section*{Appendix A: Lattice Methods}

Here we discuss the method used for the numerical evolution of the
equations of
motion. The method
employs the techniques of lattice gauge theory adapted to a
Minkowski metric
rather
than a Euclidean metric.
The formalism has been described by Moore \cite{moore96}
for
an SU(2) gauge
group and was modified to suit our case of interest,
a U(1) gauge
theory.

We approximate space-time by a Cartesian grid whose nodes
are taken to have spatial separation $a$.
Time is discretised with a fundamental time step $ha$,
$h<1$. Let us also define $l_\mu$ to be
 the lattice spacing in the
$\mu$
direction, so that
\mbox{$l_0=ha$} and \mbox{$l_i=a$}, $i=1,2,3$.
The quantity $l$ is not a Lorentz vector, and
so
raising or lowering of indices is
purely notational: $l_\mu\equiv l^\mu$. A vector of length
$l_\mu$ in the $\mu$ direction is denoted by $\hat{\mu}$.

We proceed by introducing a unit-modulus complex vector field $V:(\mu,x)
\mapsto$ U(1) such that
$V_\mu(x)$ lives on the link
between the neighbouring lattice sites $x$ and $x+\hat{\mu}$.
A scalar
field $\phi :x\mapsto {\bf C}$, with
$\phi(x)$
living on the node at $x$, completes the matter content of the model.
The vector field living on the links between the sites alters the way
the scalar field is transported
around the lattice in much the same way as a
connection in
general
relativity affects the transport of tensor quantities. The
lattice
covariant derivative
of $\phi(x)$, $\tilde{D}_\mu\phi$, is defined by
\begin{eqnarray}
\label{covderlat}
\left(\tilde{D}_\mu \phi\right)(x)&=&\frac{1}{l^\mu}
\left[V_\mu(x)\phi(x+\hat{\mu})-\phi(x)\right]~.
\end{eqnarray}
Here there is no summation over repeated indices; in the following
any summations shall be made explicit.
The covariant derivative (\ref{covderlat}) transforms in the
same way
as
the scalar field under a local U(1) transformation provided that we have
\begin{eqnarray}
\label{gauge}
\phi(x)&\longrightarrow&\Omega(x)\phi(x)~, \qquad\qquad
\Omega(x)\in\rm{U(1)}~;\nonumber\\*
V_\mu(x)&\longrightarrow&\Omega(x)V_\mu(x)\Omega^\dagger (x+\hat{\mu})~.
\end{eqnarray}
We then have the following candidate for a gauge-invariant
kinetic term of
the
scalar field,
\begin{eqnarray}
\label{Higgskin}
\sum_\mu \left(\tilde{D}_\mu \phi\right)^\dagger(x)\left(
\tilde{D}^\mu \phi\right)(x)~.
\end{eqnarray}
A gauge-invariant kinetic term for the complex vector field $V_\mu(x)$ can
be
found
by realising that the quantity
\begin{eqnarray}
V_{\mu\nu}(x)&\equiv&V_\nu(x) V_\mu(x+\hat{\nu}) V_\nu^\dagger(x+\hat{\mu})
V_\mu^\dagger (x)
\end{eqnarray}
is gauge-invariant. The relative normalisation of the scalar and
vector
kinetic
terms is determined from the continuum limit $a\to 0$
 by making the identification
$V_\mu(x)\equiv\exp(-i l_\mu e A_\mu)$. In the limit of
 small $a$ one obtains\footnote{In this Appendix we use
a sign convention for the covariant derivative which is the
opposite from that used in the rest of the paper.}
\begin{eqnarray}
\label{latcovder}
\left(\tilde{D}_\mu \phi\right)(x)&=& \partial_\mu\phi(x)
-ieA_\mu(x)\phi(x)+{\cal O}(a^2)~,\\*
{\rm Re}\,V_{\mu\nu}(x)&=&1-\frac{1}{2}l_\mu l_\nu a^2e^2
F_{\mu\nu}F_{\mu\nu}
+{\cal O}(a^4)~,\\*
F_{\mu\nu}&=&\partial_\mu A_\nu-\partial_\nu A_\mu~.
\end{eqnarray}
In terms of these quantities
we may then write down a lattice action with a discrete version
the gauge symmetry that reduces to the continuum U(1) gauge theory as
 $a\to 0$.
\begin{eqnarray}
\label{lataction}
\tilde{S}=\sum_{\stackrel{\textnormal{sites}}{\textnormal{and
links}}}ha^4
\left\{
\sum_{\mu}\left[\frac{1}{l_\mu^2}\left(
\phi^\dagger(x+\hat{\mu})V_\mu^\dagger(x)
-\phi^\dagger(x)\right)
\left(\rule{0pt}{3mm} V_\mu(x)\phi(x+\hat{\mu})
-\phi(x)\right)\right]\right.\nonumber\\*
\left.\qquad\qquad\qquad
+ \frac{2}{a^2e^2}\sum_{\mu,\nu}\frac{1}{l_\mu l_\nu}
{\rm Re} \left[V_\nu(x) V_\mu(x+\hat{\nu}) V_\nu^\dagger(x+\hat{\mu})
V_\mu^\dagger (x)\right]
-{\cal V}\left(|\phi(x)|\right)\right\}\!~.
\end{eqnarray}
Here ${\cal V}$ is the Higgs potential for the scalar field $\phi$.
The equations of
motion are derived by requiring that the lattice
analogue of functional differentiation of $\tilde{S}$ vanish
with respect to all the fields. These functional derivatives can be
calculated
using the relations
\begin{eqnarray}
\frac{\delta \phi(x)}{\delta \phi(y)}&=&\frac{\delta_{x,y}}{ha^4}~,\\*
\frac{\delta V_\mu(x)}{\delta
V_\nu(y)}&=&\delta_\mu^\nu\frac{\delta_{x,y}}%
{ha^4}~,\\*
\frac{\delta V^\dagger_\mu(x)}{\delta V_\nu(y)}&=&-V^\dagger_\mu(x)
V^\dagger_\mu(x)
\delta_\mu^\nu\frac{\delta_{x,y}}{ha^4}~.
\end{eqnarray}
Upon doing so one discovers that
the
evolution of $\phi(x)$ is not fully determined. This is a consequence
of the gauge-invariance of the system and can be remedied by making
a gauge choice analogous to the temporal gauge,
\mbox{$V_0=1$}. The equations of motion are then found to be
\begin{eqnarray}
\label{pieom}
\pi(\underline{x},t+\Delta t/2)= \pi(\underline{x},t-\Delta
t/2)\hspace*{5cm}~\nonumber\\*
  {}+ha\left\{\frac{1}{a^2}\sum_{i}\left[
V_i(\underline{x},t)\phi(\underline{x}+\hat{\imath},t)
-2\phi(\underline{x},t)
+V_i^\dagger(\underline{x}-\hat{\imath},t)\phi(\underline{x}-
\hat{\imath},t)\right]
-\frac{\partial {\cal V}}{\partial \phi^\dagger} \right\}~,
\end{eqnarray}
\begin{eqnarray}
\label{Eeom}
{\rm Im}\left[E_k(\underline{x},t+\Delta t/2)\right]&=&
{\rm Im}\left[E_k(\underline{x},t-\Delta t/2)\right]\\*
\nonumber
&+&ha\left\{\frac{2e}{a}{\rm Im}\left[\phi^\dagger(\underline{x}
+\hat{k},t)V^\dagger_k(\underline{x},t)\phi(\underline{x},t)\right]
\right.\\*
\nonumber          &~&-\frac{1}{ea^3}\sum_{i}\left( V_k(
\underline{x},t)V_i(\underline{x}+\hat{k},t)
V^\dagger_k(\underline{x}+\hat{\imath},t)V^\dagger_i(\underline{x},t)
\right.\\*
\nonumber&~&\qquad+\left.V_i(\underline{x}-\hat{\imath},t)
V_k(\underline{x},t)
V^\dagger_i(\underline{x}+\hat{k}-\hat{\imath},t)V^\dagger_k(
\underline{x}-\hat{\imath},t)\right){\left.\rule{0pt}{5mm}\right\}}~.
\end{eqnarray}
In writing this we have found it convenient to define the quantities
$\pi(x)$ and $E_i(x)$ through the relations
\begin{eqnarray}
\label{phieom}
\phi(\underline{x},t+\Delta t)&=&\phi(\underline{x},t)+ha\pi
(\underline{x},t+\Delta t/2)~,\\*
\label{Veom}
V_i(\underline{x},t+\Delta t)&=&e a^2 h E_i(\underline{x},t+\Delta
t/2)V_i
(\underline{x},t)~.
\end{eqnarray}
Eqns.~(\ref{pieom})--(\ref{Veom})
define a leap-frog algorithm where the canonical momenta, $\pi(x)$ and
$E_i(x)$, are defined on time slices half-way between
the time slices where the fields live. Note that (\ref{Eeom}) only
shows
how to evolve the imaginary part of
$E_i$. To get the full evolution we use the fact that $V_i$ has unit
modulus; Eq.~(\ref{Veom}) therefore gives
\mbox{$|E_i|=1/(e a^2 h)$}, which fixes the real part of
$E_i$ up to a sign.
To determine the sign we write \mbox{$E_k=E_k^{\rm Re}+iE_k^{\rm Im}$}
and substitute
\mbox{$V_k=\exp(-ieaA_k)$} into (\ref{Veom}), finding
to lowest order in $h$:
\begin{equation}
\exp\left(-ieaA_k-iea^2h\dot{A}_k\right)\sim ea^2h\left(E_k^{\rm
Re}+i
E_k^{\rm Im}
\right)\exp\left(-ieaA_k\right)~,
\end{equation}
i.e.,
\begin{equation}
1-iea^2h\dot{A_k}\sim ea^2h\left(E_k^{\rm Re}+iE_k^{\rm Im}\right)~.
\end{equation}
We find therefore that in the small-$h$ limit the sensible sign for
the
real part of $e E_k$ is positive.
The sign of the charge $e$ is defined by matching
Eq.~(\ref{latcovder})
to the usual covariant derivative in the continuum limit.
This completes the evolution equations. The equation found by varying
the
action with respect to $V_0$ gives the
following lattice Gauss constraint,
\begin{eqnarray}
\label{gauss}
\frac{1}{a}\sum_{i}{\rm Im}
\left[E_i(x,t+\Delta t/2)-E_i(x-\hat{\imath},t+\Delta
t/2)
\right] = \nonumber\\*
 \vspace*{2cm} 2e{\rm Im}
\left[\pi^\dagger(x,t+\Delta t/2)\phi(x,t)\right]~,
\end{eqnarray}
which is naturally satisfied because the action is invariant under
the discrete
version of the gauge symmetry, Eq.~(\ref{gauge}).

When working on the lattice one must find appropriate lattice
quantities
which reduce to the objects under study
in the continuum limit. In this work
we have been particularly interested in the variation of
the phase of the scalar field. The gauge-invariant gradient of the phase
is essentially equal to the N\"other current.
This
gives us the motivation for defining the
lattice covariant derivative of the phase
\begin{eqnarray}
\tilde{D}\theta=\frac{i}{2\phi^\dagger
\phi}\left((\tilde{D}\phi)^\dagger
\phi-\phi^\dagger\tilde{D}\phi\right)~,
\end{eqnarray}
which is clearly gauge-invariant. For the lattice version of the
Abelian
field strength we use the relation
\begin{eqnarray}
{\rm Im}\, V_{\mu\nu}&=&el_\mu l_\nu F_{\mu\nu}+{\cal O}(a^4),
\end{eqnarray}
which motivates the lattice definition of the field strength as
\begin{eqnarray}
\tilde{B}_i=\frac{1}{2ea^2}\epsilon_{ijk}\,{\rm Im}\,V_{jk}.
\end{eqnarray}
\setcounter{equation}{0}
\renewcommand{\theequation}{B.\arabic{equation}}
\section*{Appendix B: Polylogarithms}
The poly-logarithm function $\li{n}(z)$ is defined by
\begin{equation}
\label{lidef2}
\li{n}(z)=\frac{(-1)^n}{(n-1)!}\int_0^1 dx \frac{\ln^{n-1}x}{x-1/z}
=\sum_{k=1}^{\infty} \frac{z^k}{k^n}~.
\end{equation}
For real $z$ and integer $n$
the power series is convergent for $-1\leq z<1$, $n>0$ and for
$z=1$, $n>1$. The functions $\li{n}(z)$ are real-valued
for $z\leq 1$. In the case $n=2$ it is convenient to use
the functional relations
\begin{equation}
\label{funcrel1}
\li{2}(z) =\left\{\begin{array}{ll}
\displaystyle
-\li{2}(1-z) +\frac{\pi^2}{6} - \ln z \ln(1-z)~,&\frac{1}{2}<z\leq
1\\*\displaystyle
\li{2}\left(\frac{1}{1-z}\right) -\frac{\pi^2}{6} +\frac{1}{2}
\ln^2\frac{1-z}{|z|} -\frac{1}{2}\ln^2|z|~,& z<-1\\*
\displaystyle
-\li{2}\left(\frac{|z|}{1-z}\right) -\frac{1}{2}\ln^2(1-z)~,& -1\leq z
<0~.
\end{array}\right.
\end{equation}
to express $\li{2}(z)$
in terms of
functions $\li{2}(w)$  where
$0\leq w\leq 1/2$. For such $w$ the power series in Eq.~(\ref{lidef2})
converges rapidly.
For general complex $z$ we also have the following relations:
\begin{equation}
\label{minrel}
\li{2}(z) +\li{2}(1-z) = \frac{\pi^2}{6} - \ln z \ln (1-z)~,
\end{equation}
\begin{equation}
\label{recrel}
\li{2}(z) +\li{2}\left(\frac{1}{z}\right)=-\frac{\pi^2}{6}
-\frac{1}{2}
\ln^2(-z)~,
\end{equation}
{}From Eqs.~(\ref{recrel}) and (\ref{lidef2}) it follows that the
dilogarithm function $\li{2}$ has the following asymptotic
behaviour:
\begin{equation}
\label{li2asymp}
\li{2}(x)\to -\frac{1}{2}\ln^2(-x) - \frac{\pi^2}{6}~,\quad\quad x\to
-\infty~.
\end{equation}
More relations between $\li{2}$ functions
are given in Ref.~\cite{lewin}.

\end{document}